\documentclass[twocolumn,times]{aastex63}

\usepackage[english]{babel}
\usepackage{newtxtext}
\usepackage[T1]{fontenc}
\usepackage{ae, aecompl}
\usepackage{etoolbox}
\usepackage{amsmath, amssymb}
\usepackage{graphicx, mathptm, mathrsfs} 
\usepackage{times} 
\usepackage{natbib}
\usepackage{bigints}
\usepackage{color, xcolor, colortbl}
\usepackage{soul}
\usepackage{booktabs}
\usepackage{epstopdf}
\usepackage[title]{appendix}

\usepackage{scalerel}
\usepackage{tikz}
\usetikzlibrary{svg.path}

\def \apjl{Astrophys. J. Lett.}
\def \apjs{Astrophys. J. Suppl.}

\def \aap{Astron. Astrophys.}

\newcommand{\Mpch}{$h^{-1}\,\mbox{Mpc}$\,}

\defcitealias{wen2012}{WHL12}
\defcitealias{Planck2018}{Planck18}
\defcitealias{taruya2010}{TNS}
\defcitealias{dvali2000}{DGP}

\submitjournal{ApJ}

\shorttitle{BAO in the 3PCF of SDSS galaxy clusters}

\shortauthors{M. Moresco et al.}

\begin{document}

\title{C$^3$ - Cluster Clustering Cosmology \\II. First detection of the BAO peak in the three-point correlation function of galaxy clusters}

\correspondingauthor{Michele Moresco}
\email{michele.moresco@unibo.it}

\author[0000-0002-7616-7136]{Michele Moresco}
\affiliation{Dipartimento di Fisica e Astronomia ``Augusto Righi'' - Alma Mater
  Studiorum Universit\`{a} di Bologna, via Piero Gobetti 93/2, I-40129
  Bologna, Italy}
\affiliation{INAF - Osservatorio di Astrofisica e Scienza dello Spazio
  di Bologna, via Piero Gobetti 93/3, I-40129 Bologna, Italy}

\author[0000-0003-2387-1194]{Alfonso Veropalumbo}
\affiliation{Dipartimento di Fisica, Universit\`{a} degli Studi Roma
  Tre, via della Vasca Navale 84, I-00146 Rome, Italy}

\author[0000-0002-8850-0303]{Federico Marulli}
\affiliation{Dipartimento di Fisica e Astronomia ``Augusto Righi'' - Alma Mater
  Studiorum Universit\`{a} di Bologna, via Piero Gobetti 93/2, I-40129
  Bologna, Italy}
\affiliation{INAF - Osservatorio di Astrofisica e Scienza dello Spazio
  di Bologna, via Piero Gobetti 93/3, I-40129 Bologna, Italy}
\affiliation{INFN - Sezione di Bologna, viale Berti Pichat 6/2,
  I-40127 Bologna, Italy}

\author[0000-0002-3473-6716]{Lauro Moscardini}
\affiliation{Dipartimento di Fisica e Astronomia ``Augusto Righi'' - Alma Mater
  Studiorum Universit\`{a} di Bologna, via Piero Gobetti 93/2, I-40129
  Bologna, Italy}
\affiliation{INAF - Osservatorio di Astrofisica e Scienza dello Spazio
  di Bologna, via Piero Gobetti 93/3, I-40129 Bologna, Italy}
\affiliation{INFN - Sezione di Bologna, viale Berti Pichat 6/2,
  I-40127 Bologna, Italy}

\author[0000-0002-4409-5633]{Andrea Cimatti}
\affiliation{Dipartimento di Fisica e Astronomia ``Augusto Righi'' - Alma Mater
  Studiorum Universit\`{a} di Bologna, via Piero Gobetti 93/2, I-40129
  Bologna, Italy}
\affiliation{INAF - Osservatorio di Astrofisica e Scienza dello Spazio
  di Bologna, via Piero Gobetti 93/3, I-40129 Bologna, Italy}

\begin{abstract}
Third-order statistics of the cosmic density field provides a powerful cosmological probe containing synergistic information to the more commonly explored second-order statistics. Here, we exploit a spectroscopic catalog of 72,563 clusters of galaxies extracted from the Sloan Digital Sky Survey, providing the first detection of the baryon acoustic oscillations (BAO) peak in the three-point correlation function (3PCF) of galaxy clusters. We measure and analyze both the connected and the reduced 3PCF of SDSS clusters from intermediate ($r\sim10$ Mpc/h) up to large ($r\sim140$ Mpc/h) scales, exploring a variety of different configurations. From the analysis of reduced 3PCF at intermediate scales, in combination with the analysis of the two-point correlation function, we constrain both the cluster linear and non-linear bias parameters, $b_1=2.75\pm0.03$ and $b_2=1.2\pm0.5$. We analyze the measurements of the 3PCF at larger scales, comparing them with theoretical models. The data show clear evidence of the BAO peak in different configurations, which appears more visible in the reduced 3PCF rather than in the connected one. From the comparison between theoretical models considering or not the BAO peak, we obtain a quantitative estimate of this evidence, with a $\Delta \chi^2$ between 2 and 75, depending on the considered configuration.  Finally, we set up a generic framework to estimate the expected signal-to-noise ratio of the BAO peak in the 3PCF exploring different possible definitions, that can be used to forecast the most favorable configurations to be explored also in different future surveys, and applied it to the case of the Euclid mission.
\end{abstract}

\keywords{Observational cosmology -- Cosmological parameters -- Redshift surveys -- Galaxy clusters}

%%%%%%%%%%%%%%%%%%%%%%%%%%%%%%%%%%%%%%%%%%%%%%%%%%%%%%%%%%%%%%%%%%%%%%%%%%%%%%%
%%%%%%%%%%%%%%%%%%%%%%%%%%%%%%%%%%%%%%%%%%%%%%%%%%%%%%%%%%%%%%%%%%%%%%%%%%%%%%%

\section{Introduction}
Since the discovery of the accelerated expansion of the Universe
\citep{riess1998,perlmutter1999}, two main issues have been the focus
of modern cosmology: what are the main components of our Universe, and
how the Universe evolves. These questions involve the understanding of
both the geometry of our Universe and its evolution. To address these
issues many different cosmological probes have been introduced and
studied in the last twenty years; for a complete review, we refer to
the recent work by \cite{huterer2018}.

In recent years, the study of the clustering of large scale
structures has rapidly become one of the main cosmological probes,
because it retains cosmological information of the primordial Universe
in the form of peculiar matter overdensities that appear around 100 
$h^{-1}$ Mpc, as initially predicted in the seminal works of
\cite{sunyaev1970} and \cite{peebles1970}. These features are called
baryon acoustic oscillations (BAO), and can be used as {\it standard
  rulers} to constrain the expansion history of the Universe. In
Fourier space, they appear as wiggles in the power spectrum, $P(k)$, 
while in configuration space as a distinctive peak
around $r\sim$ 100 $h^{-1}$ Mpc in the two-point correlation function
(2PCF).

After the first BAO measurements by \cite{eisenstein2005} and
\cite{cole2005}, several works followed, exploring in detail the
cosmological constraining power of BAO
\citep{percival2007,blake2011,beutler2011,padmanabhan2012,anderson2014,ross2015,alam2017,bautista2020,gilmarin2020},
leading to several surveys and space missions that are currently being
developed aiming at pushing the accuracy of this probe down to the
percentage level, like Euclid \citep{euclid}, the Vera C. Rubin
Observatory LSST \citep{lsst2019}, and the Nancy Grace Roman Space
Telescope \citep{spergel2015}.

As a parallel effort to the study of two-point statistics,
increasingly more attention is being given to higher-order
correlation functions, and in particular to the three-point
statistics, namely the three-point correlation function (3PCF) in
configuration space, and its analogous in Fourier space, the
bispectrum $B(k)$. Pioneering studies on the 3PCF can be traced back to 
the late seventies \citep{peebles1975,groth1977}. Historically, the 3PCF 
has been mainly exploited to characterize extra-galactic samples in 
terms of their spatial distribution as a function of properties such
as luminosity or stellar mass \citep{fry1994, frieman1994, jing1995,
  jing1998, jing2004, kayo2004, gaztanaga2005, nichol2006,
  kulkarni2007, mcbride2011, marin2011, guo2014, guo2015, guo2016,
  moresco2017}, while in recent years it has been used also to place
constraints on the evolution of some parameters, such as the sample
bias or the matter power spectrum normalization, $\sigma_8$, as a
function of redshift \citep{marin2013, moresco2014, hoffmann2015}. 
For more extensive reviews on the topic, we refer to
\citet{bernardeau2002}, \cite{takada2003}, and \cite{desjacques2018}.

Studies on the exploitation of the 3PCF for cosmological purposes, and
in particular to detect and constrain BAO, are, instead, scarcer. The
first detection of the BAO peak in the 3PCF is provided by
\cite{gaztanaga2009} from the analysis of Sloan Digital Sky Survey
(SDSS) Data Release 7 (DR7) Luminous Red Galaxies. More recently,
\cite{slepian2017, slepian2017b} obtained a clear detection of the BAO
feature in the 3PCF of SDSS-DR12 CMASS sample, while
\cite{decarvalho2020} measured the BAO feature in the angular 3PCF of
SDSS-DR12 quasars.

In this work, we present the first significant detection of the
BAO peak in the 3PCF of galaxy clusters. Clusters of galaxies
represent the most massive virialized systems in our Universe, with
masses $\log(M/M_{\odot})\gtrsim14$, and comprising up to thousands of
galaxies. These systems are extremely important laboratories both from
astrophysical and cosmological points of view; in this latter context,
they have been used to study the properties of dark energy, since
their number density can directly provide constraints on the
underlying cosmology \citep[e.g.][]{vikhlinin2009, pacaud2018,
  costanzi2019, lesci2020}. This topic has been extensively studied
both in simulations and observations, so that the cluster number
counts currently represent a key additional cosmological probe
\citep[for a detailed review on this topic, we refer to][]{allen2011}. 
The properties of galaxy clusters make them ideal to be studied also 
from a clustering perspective. 
Being extremely massive systems, they are characterized
by a large bias, and therefore the clustering signal is stronger
compared to a galaxy sample with comparable, or also larger
densities. Moreover, the effects of non-linear dynamics have a smaller
impact on cluster than on field galaxies, with a resulting smaller
systematic effect due to redshift-space distortions (RSD)
\citep{valageas2012, marulli2017}. For these reasons, galaxy clusters
have been extensively analyzed in terms of two-point statistics
\citep[e.g.][]{estrada2009, hutsi2010, hong2012, veropalumbo2014,
  veropalumbo2016, sereno2015, hong2016, marulli2018, marulli2020,
  nanni2020}, while the analyses on higher-orders are significantly
less developed. 

This work is part of a project called ``Cluster Clustering Cosmology 
($C^3$)'' where we exploit the constraining power of cluster clustering. 
In Paper I \citep{marulli2020} we analyze the RSD of galaxy clusters, 
providing a new strong constraint on the product between the linear 
growth rate of  cosmic structures $f$ and the amplitude of linear matter 
density  fluctuations quantified at 8 \Mpch $\sigma_{8}$ at an effective 
redshift  $z\sim0.3$, $f\sigma_{8}=0.46\pm0.03$. In Paper III 
\citep{veropalumbo2020}, we explore the combination of 2PCF
and 3PCF in a joint analysis of RSD and BAO.

This paper is organized as follows. In Sect.~\ref{sec:methods} we give
an overview of the methods adopted in this analysis, presenting the
estimators used to measure the 3PCF, the models considered, and the
code developed to perform the measurements and the statistical
analysis. In Sect.~\ref{sec:data} we present the cluster catalog used
in this analysis. In Sect.~\ref{sec:analysis} we discuss the 3PCF
measurements performed, how they have been analyzed to extract
information on the bias parameters of the sample, and the advantage 
of using clusters as opposed to normal galaxies as tracers. Then in 
Sect.~\ref{sec:BAOdet} we focus on the BAO peak in the measured 3PCF, 
quantifying the statistical level of the detection. 
In Sect.~\ref{sec:BAO} we introduce a framework to
forecast the expected signal-to-noise ratio (SNR) of the BAO peak in
the 3PCF, focusing on two possible methods, one model-dependent and
one model-independent, to identify the peak, and apply them to
provide forecasts for the Euclid mission, discussing the constraining 
potential of the 3PCF as a function of methods and configurations. 
Finally, in Sect.~\ref{sec:concl}, we draw our conclusions.

Throughout this paper, we assume a Planck18 cosmology \citep{planck2018} 
when otherwise explicitly stated, with $\Omega_{\rm m}=0.3153$, 
$\Omega_{\rm DE}=0.684607$, and $\rm H_0= 67.36$ km/s/Mpc.

%%%%%%%%%%%%%%%%%%%%%%%%%%%%%%%%%%%%%%%%%%%%%%%%%%%%%%%%%%%%%%%%%%%%%%%%%%%%%%%
%%%%%%%%%%%%%%%%%%%%%%%%%%%%%%%%%%%%%%%%%%%%%%%%%%%%%%%%%%%%%%%%%%%%%%%%%%%%%%%

\section{Methods}
\label{sec:methods}
\subsection{Definitions}
Given a generic spatial distribution of objects, the 3PCF is defined
as the statistical function that provides the probability of finding
triplets of objects at corresponding comoving separations $r_{12}$,
$r_{13}$, and $r_{23}$ \citep{peebles1980}. Specifically, the 3PCF 
$\zeta$ can be defined implicitly as follows:
\begin{multline}
d{\rm
  P}=\bar{n}^{3}[1+\xi(r_{12})+\xi(r_{13})+\xi(r_{23})+\zeta(r_{12},r_{13},r_{23})]
\\ dV_{1}dV_{2}dV_{3} \, ,
\end{multline}
where $\bar{n}$ is the average density of the objects, $V_i$ the
comoving volumes at $\overrightarrow{r_{i}}$, and $\xi$ is the 2PCF of
the sample at comoving separations ${r_{ij}}$.

While several estimators have been proposed to compute the 2PCF
\citep[e.g.][]{peebles1974, hewett1982, davis1983, landy1993,
  hamilton1993, keihanen2019}, fewer studies have been performed so
far for the 3PCF \citep[e.g.,][]{jing1998, szapudi1998,
  sosa2020}.  Here we will exploit the widely used formulation
proposed by \cite{szapudi1998}, that provides an unbiased and with
minimal variance estimator based on the counting of triplets between
the input catalog (that we will label as data catalog, $D$), and a
corresponding random catalog (that we will label as $R$), constructed
to reproduce the geometric distribution of the input catalog, but with
zero clustering. Specifically, the \cite{szapudi1998} estimator is
constructed as follows:
\begin{equation}
\zeta(r_{12},r_{13},r_{23})=\frac{DDD-3DDR+3DRR-RRR}{RRR} \, ,
\end{equation}
where $DDD$, $RRR$, $DDR$, and $DRR$ are the triplets having comoving 
separations $(r_{12},r_{13},r_{23})$ in the data catalog, in the random 
catalog, and in the mixed data-random catalogs, normalized by $N^3_D/6$, 
$N^3_R/6$, $N^2_D N_R/2$, and $N_D N^2_R/2$, respectively.

With a similar approach, \cite{landy1993} have defined an estimator for 
the 2PCF, $\xi(r)$, based on pair counting in the data and random 
catalogs:
\begin{equation}
\xi(r)=\frac{DD-2DR+RR}{RR} \, ,
\label{eq:2PCF}
\end{equation}
where $DD$, $RR$, and $DR$ are the pairs having a comoving separation 
$r$ in the data catalog, in the random catalog, and in the mixed one, 
normalized by $N^2_D/2$, $N^2_R/2$, and $N_D N_R$, respectively.

Since it can be demonstrated that in hierarchical scenarios the
connected 3PCF, $\zeta$, is proportional to $\xi^{2}$
\citep{peebles1975}, a natural quantity that can be derived is the
reduced 3PCF, $Q$ \citep{groth1977}, defined as:
\begin{equation}
Q(r_{12},r_{13},r_{23}) \equiv
\frac{\zeta(r_{12},r_{13},r_{23})}{\xi(r_{12})\xi(r_{13})+\xi(r_{13})\xi(r_{23})+\xi(r_{23})\xi(r_{13})}
\, .
\label{eq:3PCF}
\end{equation}
This new quantity is particularly convenient because it is
characterized by a smaller variation of its values as a function of
scales. Its values are, by definition, close to unity at all
scales, differently from $\zeta$ that may vary by over three orders of
magnitude from small to intermediate scales. A further advantage is
that it does not depend on $\sigma_{8}$ (unlike the 2PCF or the
connected 3PCF), but only on the bias parameters (see
Sect.~\ref{sec:models}).

Several parameterizations of triangle shapes have been proposed in the
literature to study the 3PCF. Usually, they involve the definition of
two sides of the triangle, exploring the dependence of the 3PCF on the
third side \citep{jing1995, gaztanaga2005, nichol2006, kulkarni2007,
  guo2014}. Alternatively, it might be convenient to fix all sides of
the triangles (typically in an equilateral configuration) and to
analyze the 3PCF at increasing scales \citep[see e.g.][]{wang2004,
  fosalba2005, marin2008}, or to follow an approach similar to the one
adopted for the bispectrum and to study the 3PCF at all scales \citep[see
  e.g.][]{veropalumbo2020}. Here, we follow the parameterization
proposed by \citet{marin2011}, in which the first two sides of the
triangle are fixed, one as a function of the other $r_{13} = u\times
r_{12}$, and the 3PCF is calculated either as a function of the third
side, or of the angle $\theta$ between the two sides:
\begin{equation}
  \left\{
    \begin{array}{l}
      r_{12}\\
      r_{13}= u\cdot r_{12}\\
      r_{23}\equiv r_{12}\cdot\sqrt{1+u^{2}-2\cdot u \cdot \cos\theta}\, . \nonumber 
    \end{array}
    \right.
\end{equation}
In this approach, configurations with either $\theta\sim0$ or
$\theta\sim\pi$ will represent the elongated configurations, while the
ones with $\theta\sim\pi/2$ represent the isosceles configurations.

\subsection{The models}
\label{sec:models}
The real-space overdensity of galaxy clusters, $\delta_c$, can be 
expressed with a perturbative expansion as a function of the dark-matter 
one, $\delta_m$, as follows \citep{fry1993}:
\begin{equation}
\delta_c\approx b_{1}\delta_m+\frac{b_{2}}{2}\left(\delta^2_m-<\delta^2_m>\right) \, .
\label{eq:deltag}
\end{equation}
This equation, known as the local bias approximation, relates the clusters and dark matter overdensities as a function of two parameters, the linear bias $b_1$ and the non-linear bias $b_2$, neglecting a possible tidal tensor bias contribution.

Using this approximation, it is possible to express also the 2PCF and
the 3PCF of clusters as a function of the dark-matter one as follows:
\begin{eqnarray}
\xi_c &=& \left(\frac{\sigma_8}{\sigma_{8, fid}}\right)^2b_1\xi_m \nonumber \\
\zeta_c &=& \left(\frac{\sigma_8}{\sigma_{8,
    fid}}\right)^4\left[b_1^3\zeta_m+b_2b_1^2\left(\xi_m(r_{12})\xi_m(r_{12})\right)+perm\right] \nonumber \\
Q_c &=& \frac{1}{b_1}\left(Q_m+\frac{b_1}{b_2}\right) \, ,
\label{eq:Q_mod}
\end{eqnarray}
where $\sigma_{8,fid}$ is the amplitude of linear matter density 
fluctuations quantified at 8 \Mpch for the fiducial model.

As it can be inferred from the previous equations, the effect of the bias
parameters in both $\xi$ and $\zeta$ is degenerate with $\sigma_8$,
differently from the reduced 3PCF case. More recently, \cite{bel2015}
proposed also an extension of Eq.~\eqref{eq:Q_mod} which includes also
the tidal tensor bias $g_2$:
\begin{equation}
Q_c=\frac{1}{b_1}\left(Q_m+\frac{b_1}{b_2}+g_2\;Q_{non-loc}\right)\, ,
\label{eq:Q_mod_nl}
\end{equation}
where $Q_{non-loc}$ is the non-local contribution to the 3PCF.

\subsection{The code}
The theoretical and observational data computed in this analysis have
been obtained using the \texttt{CosmoBolognaLib} suite
\citep[hereafter \texttt{CBL},][] {CBL}, which is a public C++ and
Python library for cosmological measurements.  Originally, the
\texttt{CBL} library provided the basic functions to perform
measurements of the 2PCF and 3PCF in a variety of possible
configurations, to model the 2PCF and perform cosmological
computations. The core function of the library, which allows to
significantly reduce the computation time for both of the correlation
functions, is a chain-mesh (also known as linked-list) approach, which
indicizes the particles depending on their spatial position, then
looping only on the particles that actually contribute to the counts
of pairs and triplets for the considered separations.

Since then, the library has been significantly expanded with
additional functionalities. Among them, e.g., significant work on
void detection and modelization has been done, including new functions
to detect, count, and model cosmic void statistics \citep{ronconi2017,
  contarini2019, ronconi2019}. Moreover, new modules have been added
for improved 2PCF modeling \citep{garcia2020}.

For the purpose of this analysis, the \texttt{CBL} library has been
significantly updated, with respect to the previous releases, to
include new features related to both the analysis and theoretical
modelization of the 3PCF. In particular, the chain-mesh approach has
been improved with a new method to save triplets during their
counting that allows a fast computation of errors based on either
jack-knife (JK) or bootstrap (BS) approaches.  More recently,
\cite{slepian2015} proposed an alternative method to calculate the
3PCF, based on a spherical harmonics expansion that skips the direct
estimation of the number of triplets in a given configuration, thus
significantly reducing the computational time, at the price of a smaller
accuracy for the isosceles configurations\footnote{Formally, the same
  accuracy could be reached in every configuration. However, for the
  isosceles ones it would be necessary an expansion up to high
  multipoles to recover the shape of the 3PCF with sufficient
  accuracy.}. This method has been added also in the \texttt{CBL}
library, so that the user can choose if estimating the 3PCF either
with the direct-counting method or with the
spherical-harmonics-decomposition approach.

Two different 3PCF theoretical models have been also implemented, both
for connected and reduced 3PCFs, taken from \cite{barriga2002} and
\cite{slepian2015}. These models are obtained as the anti-Fourier
transform of the real-space tree-level halo bispectrum. For symmetry
reasons, this transformation reduces to a combination of 1D integrals of
the linear power spectrum.  \citet{slepian2017b} used a similar
approach to derive the three-point correlation function model from the
tree-level redshift-space bispectrum.
A theoretical estimate of the covariance matrix for $\zeta$, which
assumes a Gaussian Random Field density and a boundary-free survey,
has also been added to the library, as proposed by
\cite{slepian2015}. Finally, different models to estimate bias
parameters have also been included (see Sect. \ref{sec:analysis}).
All these new functions are fully documented in the official web-page
of the project.\footnote{The code is publicly available at
  \url{https://gitlab.com/federicomarulli/CosmoBolognaLib}, and its
  full documentation can be found at
  \url{http://federicomarulli.github.io/CosmoBolognaLib/Doc/html/index.html}.}

%%%%%%%%%%%%%%%%%%%%%%%%%%%%%%%%%%%%%%%%%%%%%%%%%%%%%%%%%%%%%%%%%%%%%%%%%%%%%%%
%%%%%%%%%%%%%%%%%%%%%%%%%%%%%%%%%%%%%%%%%%%%%%%%%%%%%%%%%%%%%%%%%%%%%%%%%%%%%%%

\section{Data}
\label{sec:data}
The galaxy cluster sample used in this analysis has been obtained with
the same procedure discussed by \cite{veropalumbo2016}, that is from a
cross-correlation between the photometric cluster catalog provided by
\cite{wen2012} and the spectroscopic redshifts available from the SDSS
Data Release 12 (DR12), considering both the Main Galaxy Sample (MGS)
and BOSS surveys.

The parent catalog comprises 132,684 clusters extracted from SDSS-III
between $0.05\leq z<0.8$; these clusters have been selected from the
photometric galaxy catalog, over $\sim14,000$ square degrees of the
SDSS, with an improved version of the Friend-of-Friend algorithm
\citep{huchra1982}. The obtained sample is characterized by a richness
$R_{L*}>12$, and more than eight member candidates within $r_{200}$,
with typical masses larger than $M_{200}/M_{\odot}>6\times10^{13}$,
where $r_{200}$ and $M_{200}$ are, respectively, the radius within
which the mean density of the cluster is 200 times larger than the
critical density of the Universe at the cluster redshift, and the mass enclosed within $r_{200}$.

This catalog has been cross-correlated with the spectroscopic redshift
information available in SDSS-DR12
\citep{dawson2013,alam2015,alam2017}, with a significantly increased
accuracy in the determination of the position of the galaxies with respect
to the photometric ones ($\sigma_z/z\sim0.001$ instead of
$\sigma_z/z\sim0.03$), which is proven to be fundamental for clustering
measurements. This procedure assigns to each cluster the spectroscopic
redshift of its brightest central galaxy (BCG), when available.

\begin{figure}
  \includegraphics[width=0.5\textwidth]{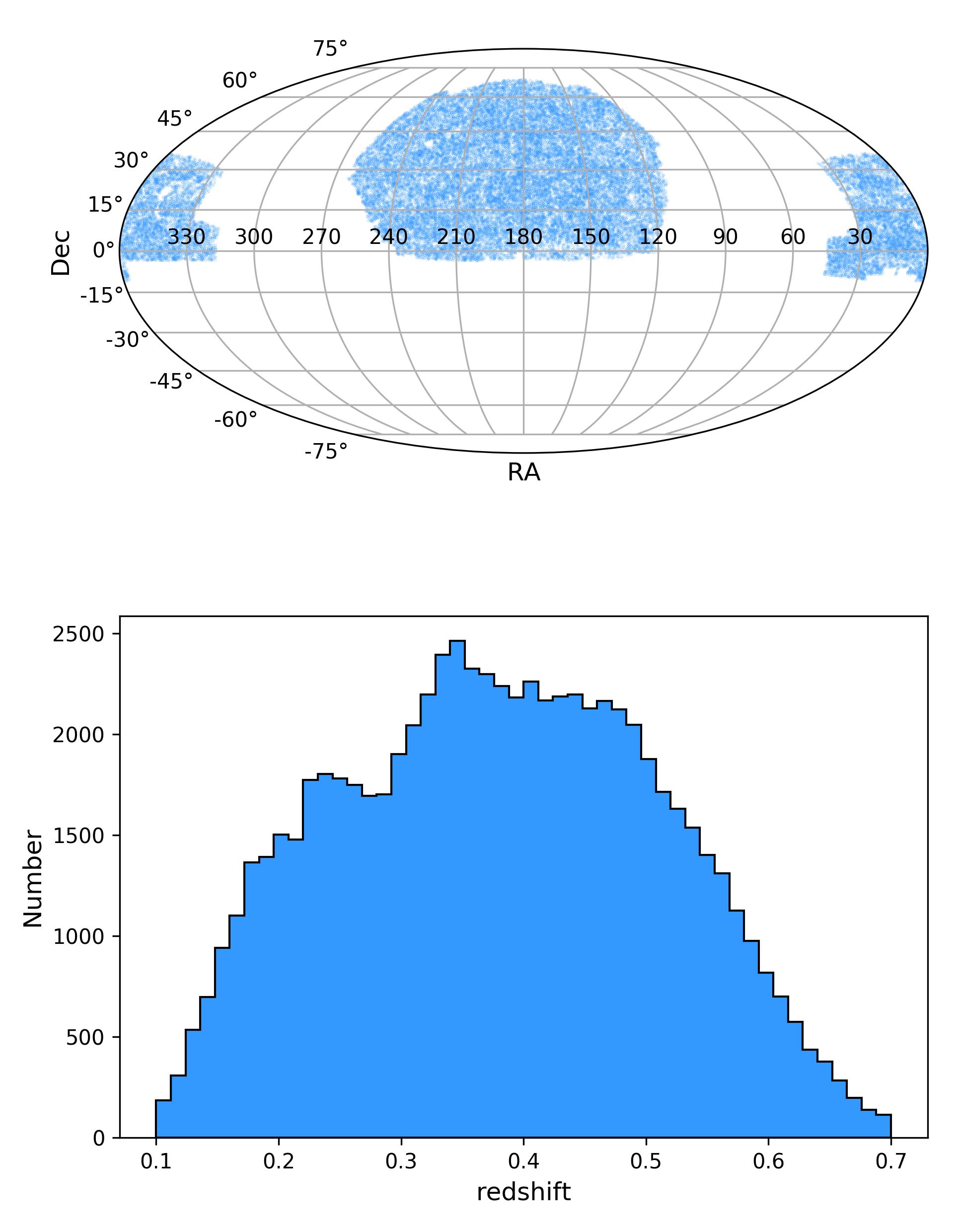}
  \caption{Angular position and redshift distribution of the galaxy 
  cluster sample.}
  \label{fig:sample}
\end{figure}

The final sample comprises 72,563 galaxy clusters in the redshift range $0.1<z<0.7$. The mean redshift is $<z>=0.38$.
Their angular position and redshift distributions are shown in Fig. \ref{fig:sample}. 

An associated random catalog has been built following the same
approach used in the BOSS collaboration \citep{reid2016}. The
\texttt{MANGLE} code \citep{swanson2008} has been used to track the
coordinates of our sample following the spectroscopic tiles
of the survey. A Gaussian kernel of $\sigma_z=0.02$ has then been used
to smooth the redshift distribution. We verified that this assumption
does not have a significant impact on the final results. 
The random catalog has been extracted with a number of objects 
$N_R=50\times N_D$, to significantly reduce the shot-noise contribution 
to the final error budget. For more details, we refer to 
\cite{veropalumbo2016} and \cite{marulli2020}.

%%%%%%%%%%%%%%%%%%%%%%%%%%%%%%%%%%%%%%%%%%%%%%%%%%%%%%%%%%%%%%%%%%%%%%%%%%%%%%%
%%%%%%%%%%%%%%%%%%%%%%%%%%%%%%%%%%%%%%%%%%%%%%%%%%%%%%%%%%%%%%%%%%%%%%%%%%%%%%%

\begin{figure*}
  \includegraphics[width=0.99\textwidth]{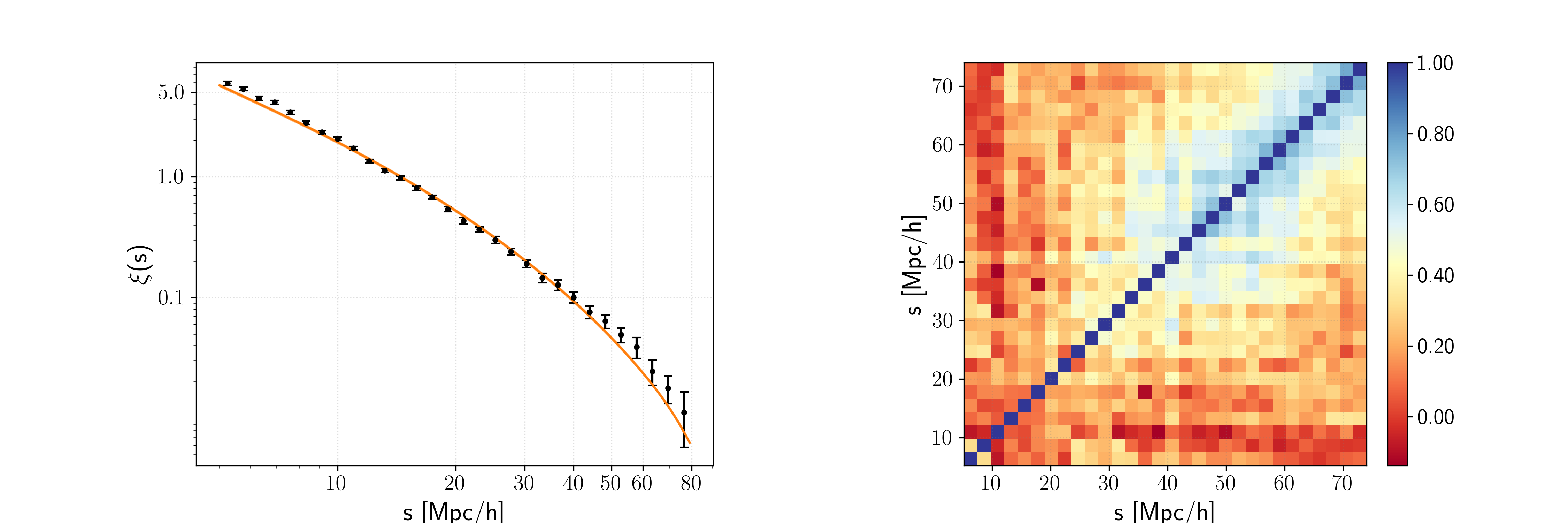}
  \caption{Monopole of the 2PCF. Left panel: the measurements compared
    to the best-fit model (orange line and shaded area indicating the
    68\% confidence level). Right panel: the corresponding correlation
    matrix.}
  \label{fig:2pcf}
\end{figure*}

%%%%%%%%%%%%%%%%%%%%%%%%%%%%%%%%%%%%%%%%%%%%%%%%%%%%%%%%%%%%%%%%%%%%%%%%%%%%%%%
%%%%%%%%%%%%%%%%%%%%%%%%%%%%%%%%%%%%%%%%%%%%%%%%%%%%%%%%%%%%%%%%%%%%%%%%%%%%%%%

\section{Analysis}
\label{sec:analysis}
In this Section, we present the analysis performed to measure the 3PCF of galaxy clusters, the constraints obtained on the bias parameters, and a comparison with results obtained in the literature.

\subsection{The measurements}
\label{sec:measzeta}
For the purpose of our analysis, we measure the 3PCF at several 
different scales, and in different configurations. 

As a first step, we assess the bias of the population, thus estimating
the 3PCF at intermediate scales. We do not consider the small scales 
($r<$10 $h^{-1}$ Mpc) that are not accurately reproduced by models due to the non-linear dynamics. 
In this first analysis, we do not consider the BAO scales as
well, so that our results will not be affected by the BAO model
considered. We choose three different configurations, with $r_{12}=10$
$h^{-1}$ Mpc and $r_{13}=$ 30, 50, and 70 $h^{-1}$ Mpc,
respectively. For each case, we consider 15 angular bins in $\theta$,
defining a third side of the triangle spanning $20<r_{23}<40$ $h^{-1}$
Mpc, $40<r_{23}<60$ $h^{-1}$ Mpc, and $60<r_{23}<80$ $h^{-1}$ Mpc,
respectively. In this way, we consider three configurations with a
non-overlapping third side, and adopt a 5\% tolerance on the $r_{12}$
and $r_{13}$ sides of the triangle, namely $\Delta
r_{ij}/r_{ij}=0.05$. With these settings, we measure both the
connected, $\zeta$, and the reduced, $Q$, 3PCFs. We explored also
other different configurations (not shown here), finding that the main results of the
analysis do not change appreciably.

Then to detect the BAO peak, we focus on three different
configurations where the BAO peak is included in the third triangle
side $r_{23}$, namely $(r_{12}, r_{13})=$ (20, 105), (40, 100), and
(50, 90) $h^{-1}$ Mpc. The reason behind the choice of these
configurations is that they are representative of different regimes at
which we expect the BAO signal to be statistically significant, as
will be discussed in Sect.~\ref{sec:BAOdet}. Also in this case,
we measure both $\zeta$ and $Q$.

The error covariance matrices have been estimated with a JK approach,
considering 100 sub-volumes. We verified that changing this value to a
higher sampling has a negligible impact on the resulting errors.
Therefore, in the following analysis, we will adopt this choice which
allows us a fast enough numerical computation.

Finally, we also measure the monopole of the 2PCF on the same
catalog. We use the \cite{landy1993} given by Eq.~\eqref{eq:2PCF},
with 30 logarithmically-spaced bins in the range $5<r$[$h^{-1}$
  Mpc]$<80$. These data will be helpful to provide further additional
information on the bias of our sample in our analysis, as discussed in
Sect.~\ref{sec:measbias}.

The measurements for the 2PCF and the 3PCF are presented in 
Figs.~\ref{fig:2pcf} and \ref{fig:scales_b1b2}. All of 
them have been performed assuming a Planck18 cosmology to convert 
observed coordinates into comoving ones. We also tested the results 
assuming a Planck15 \citep{planck2016} cosmology, finding differences 
smaller than 1-$\sigma$.

\subsection{Estimating the bias}
\label{sec:measbias}

\begin{figure}[ht!]
  \includegraphics[width=0.48\textwidth]{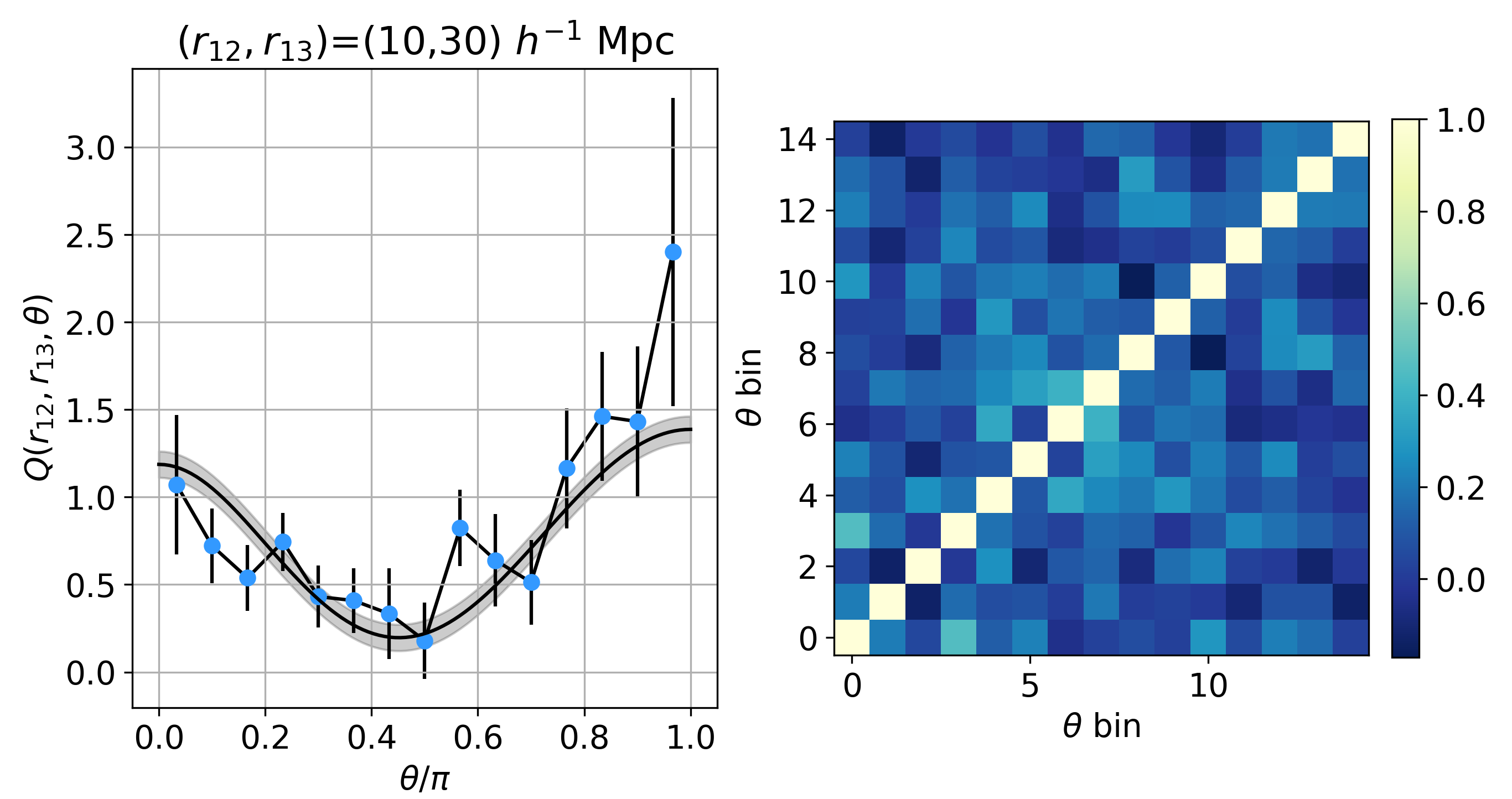}
  \includegraphics[width=0.48\textwidth]{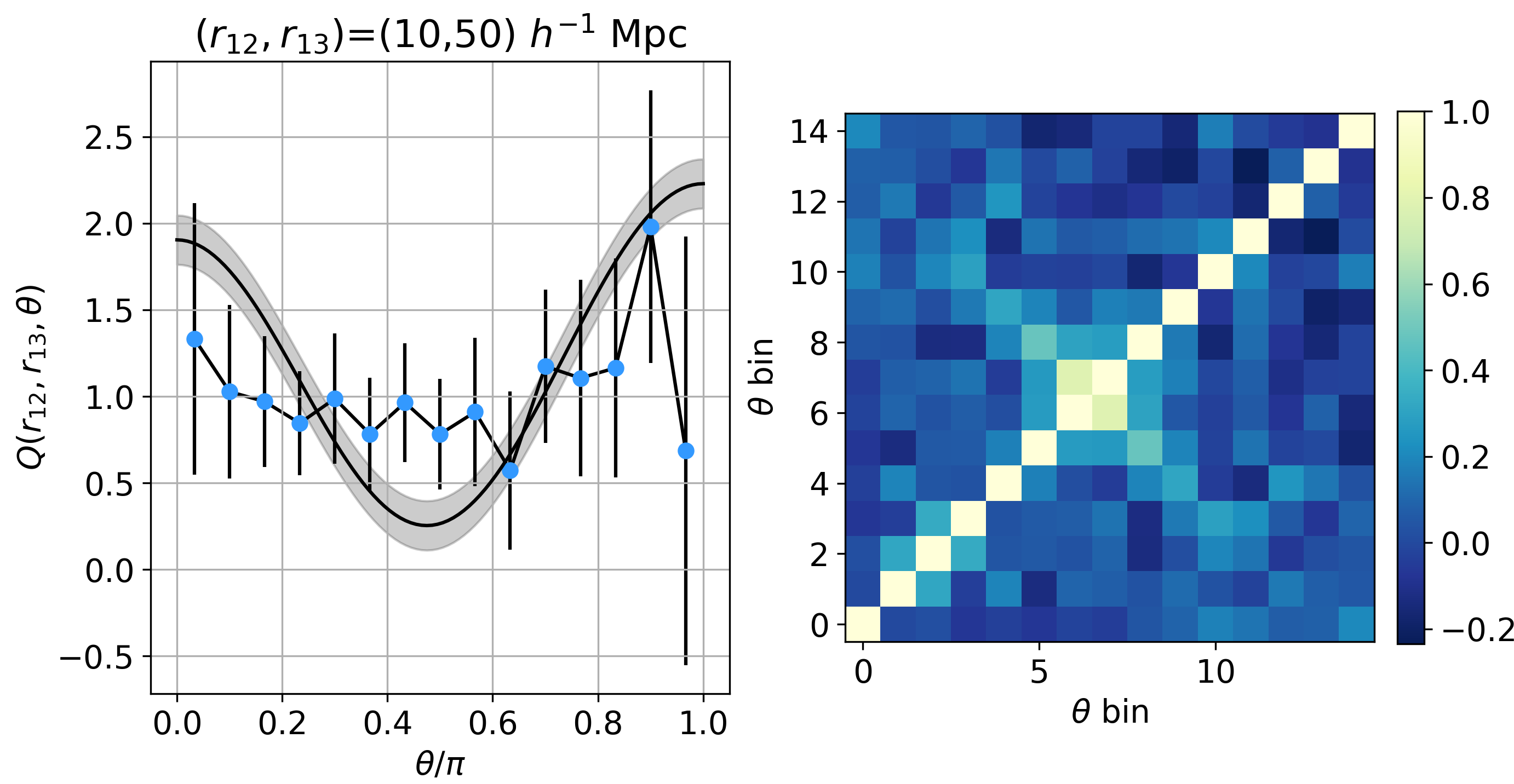}
  \includegraphics[width=0.48\textwidth]{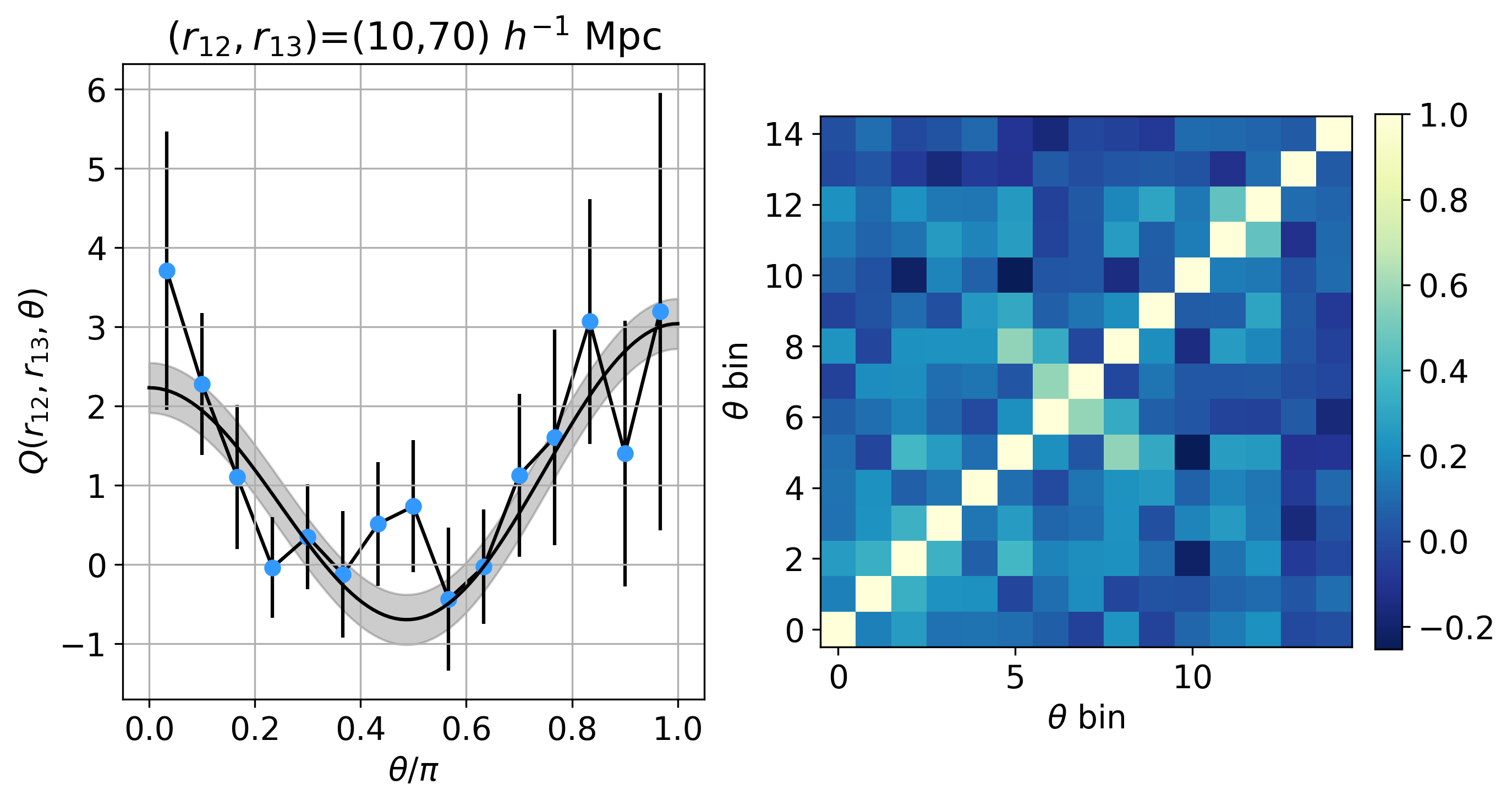}
  \caption{3PCF measurements for SDSS galaxy clusters at intermediate
    scales. The considered configurations have been chosen to have the
    first side $r_{12}=10$ $h^{-1}$ Mpc and the second side $r_{13}=$
    30, 50, and 70 $h^{-1}$ Mpc, from top to bottom. As a result, the
    third side spans the ranges $20<r_{23}$[$h^{-1}$ Mpc]$<40$,
    $40<r_{23}$[$h^{-1}$ Mpc]$<60$, and $60<r_{23}$[$h^{-1}$
      Mpc]$<80$, respectively. The left panels show the reduced 3PCF
    $Q$, while the right ones the corresponding correlation
    matrices. The best-fit 3PCF models are shown as a black curve in
    the left panels, with their associated 1-$\sigma$ uncertainties as
    shaded grey area.}
    \label{fig:scales_b1b2}
\end{figure}

\begin{figure}[ht!]
  \includegraphics[width=0.48\textwidth]{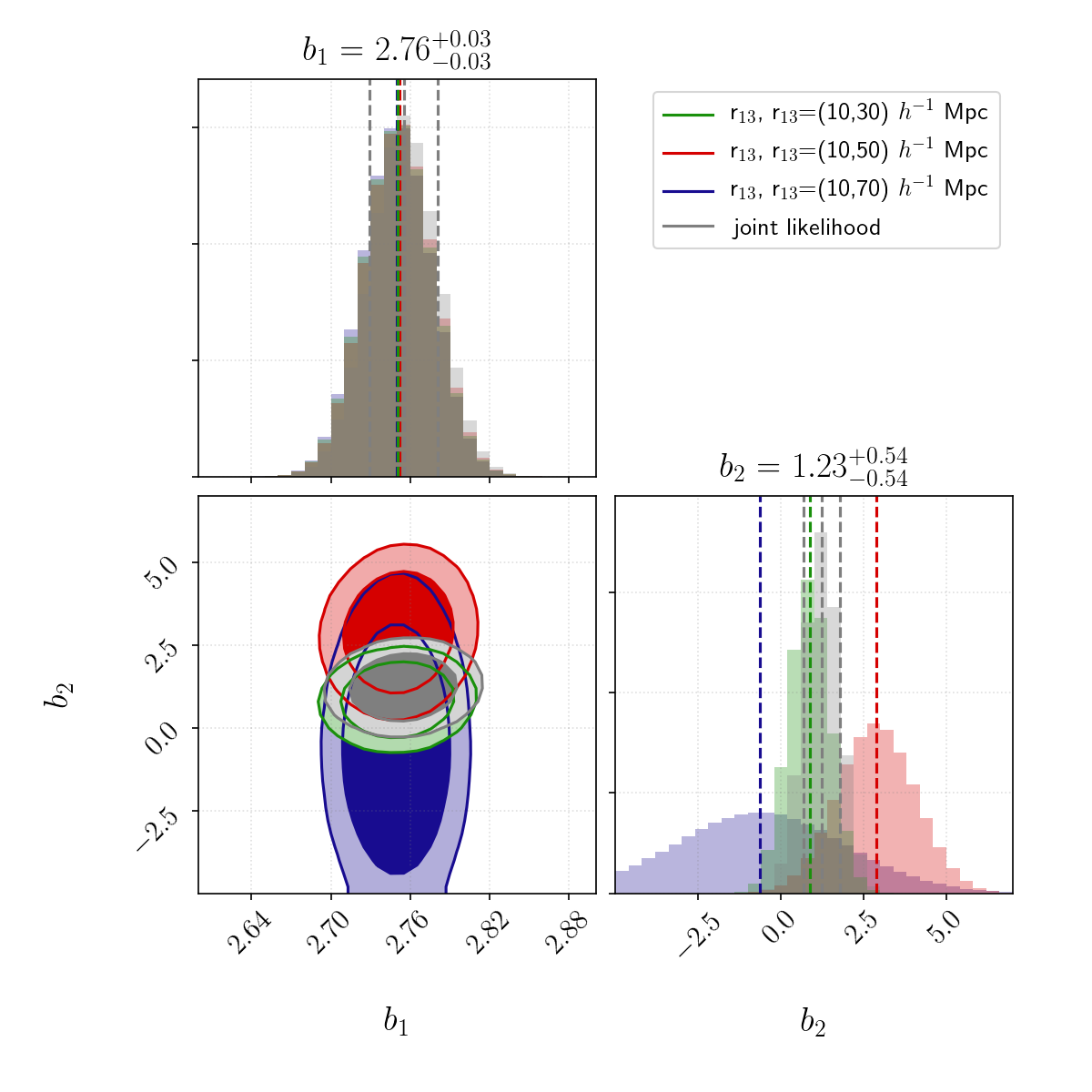}
  \caption{Constraints on $b_1$ and $b_2$ from the configurations
    shown in Fig.~\ref{fig:scales_b1b2} (shown in different colors) and
    from the joint analysis of all scales (in gray).}
  \label{fig:constraintsb1b2}
\end{figure}

To constrain the bias parameters of the selected galaxy cluster
sample, we fit the reduced 3PCF, $Q(\theta)$, which provides the
advantages discussed in Sect.~\ref{sec:models}, that is to break the
degeneracy between bias parameters and $\sigma_8$ \citep[see
  e.g.][]{kayo2004,zheng2004,guo2009,marin2011,mcbride2011,guo2014,guo2016}.

As presented above, different models can be explored to fit the 3PCF,
including bias parameters at different levels. We investigated the
possibility of fitting our data with both Eqs.~\eqref{eq:Q_mod} and
\eqref{eq:Q_mod_nl}. Due to the statistics of the current sample, we 
verified that the SNR is too low for a 3-parameters fit. Therefore we
excluded the possibility of constraining the tidal parameter $g_2$, and
in the following analysis we will consider only the model provided by
Eq.~\eqref{eq:Q_mod}.

Unfortunately, the degeneracy between $b_1$ and $b_2$ in the fit of
the reduced 3PCF does not allow us to obtain constraints enough 
stringent, even with two parameters only. We, therefore, decided to 
include in the analysis the additional information coming from the 2PCF, 
to constrain  the bias parameter $b_1$. Specifically, we model the 
monopole of the 2PCF in redshift space, $\xi(s)$, as follows 
\citep{kaiser1987, hamilton1992}:
\begin{equation}
  \xi(s) = \left[ (b_1\sigma_8)^2 + \frac{2}{3} f\sigma_8
    \cdot b_1\sigma_8 + \frac{1}{5}(f\sigma_8)^2 \right]
  \frac{\xi_{\rm DM}(r)}{\sigma_8^2} \; ,
  \label{eq:xi0} 
\end{equation}
where $f \simeq \Omega_{\rm M}^{0.545}(z)$ is the linear growth rate of cosmic structures,
and $\xi_{\rm DM}(r, z)$ is the real-space dark matter 2PCF estimated
by Fourier transforming the linear matter power spectrum, computed
with {\small CAMB} \citep{lewis2002}.  We fit the 2PCF in the range
$10<s $[$h^{-1}$ Mpc]$<60$, considering its associated covariance
matrix, and assuming a flat prior on $b_1$ between $0$ and $4$. We
limited our analysis in this range because, while the correlation
matrix is almost diagonal below 60 $h^{-1}$ Mpc, it presents a
significant correlation between different bins above, as can be seen
in Fig.~\ref{fig:2pcf}. However, we checked that changing $s_{min}$
down to 5 $h^{-1}$ Mpc and $s_{max}$ up to 80 $h^{-1}$ Mpc, the
obtained constraints on the bias show a negligible variation, below
the 1\% level. From the fit of the 2PCF monopole, we obtain
$b_1=2.75\pm0.03$. \cite{veropalumbo2016} measured and modeled the
2PCF of this same sample, but splitting it into three subsamples
depending on the survey they were belonging to (SDSS-MGS, BOSS-LOWZ,
and BOSS-CMASS). We find that our result is fully compatible with
theirs once averaged, with a difference smaller than 1-$\sigma$ 
(taking into account the different cosmological model assumed). In 
this analysis, we decided not to divide our sample into smaller 
sub-samples since we verified that it would lower the statistics, 
increasing the shot-noise excessively.

We use this bias measurement as a Gaussian prior in our 3PCF analysis,
when exploiting the model given by Eq.~\eqref{eq:Q_mod} to place
constraints on the non-linear bias parameter, $b_2$, for which we
assume a flat prior in the range $-20\leq b_2\leq20$.

\begin{table}[b!]
\centering
\caption{Bias parameter estimates for the SDSS galaxy cluster
  catalog. The linear bias is estimated from the 2PCF (assuming
  Planck18 cosmology), while the non-linear bias from the 3PCF
  (assuming Planck18 cosmology and a Gaussian prior on $b_1$ from the
  2PCF). The constraints for the non-linear bias are shown at
  different scales. The last line shows the result from the combined
  fit at all scales.}
\label{tab:tab1}
\begin{tabular}{lcccc}
\hline
method & scales [$h^{-1}$ Mpc] & $\rm b_1$ & $\rm b_2$ & $\chi^2$/d.o.f\\
\hline
\hline
2PCF & $10<s<60$ & $2.75\pm0.03$ & -- & 1.3\\
\hline
\hline
3PCF & $20<r_{23}<40$ & -- & $0.9\pm0.6$ & 1.1\\
3PCF & $40<r_{23}<60$ & -- & $2.9\pm1.2$ & 1.2\\
3PCF & $60<r_{23}<80$ & -- & $-0.6\pm2.6$ & 0.7\\
\hline
3PCF & joint scales & -- & $1.2\pm0.5$ &\\
\hline
\end{tabular}
\label{tab:b2}
\end{table}

We fit $Q(r_{12},r_{13},\theta)$ at the different scales presented in
Sect.~\ref{sec:measzeta}, considering the associated covariances. The
results are shown in Figs. \ref{fig:scales_b1b2} and
\ref{fig:constraintsb1b2}, and summarized in Tab.\ref{tab:b2}. We find
an increasing precision on the $b_2$ constraint with decreasing scale,
that can be expected as the errors on $Q$ are smaller at smaller
scales, due to the higher number of triplets that can be found in
these configurations. In all cases, the best-fit model well reproduces
the data, except in the intermediate bin, with 
($r_{12},r_{13})$=(10,50) $h^{-1}$ Mpc, where a larger discrepancy is 
observed. This effect is due to the fact that the prior on $b_1$ 
fixes the shape of the model,  while the parameter $b_2$ (left free 
in the fit) controls its  normalization (see Eq.~\ref{eq:Q_mod}). In 
all cases, however,  the reduced $\chi^2$ is between 0.7 and 1.2. 
Given that we chose non-overlapping configurations in $r_{23}$, we 
also performed a joint fit to the three scales. The results are 
presented in Fig.~\ref{fig:constraintsb1b2}, showing the 68\% and 
95\% confidence  contours for $b_1-b_2$, and their marginalized 
distribution. The  combined analysis has been performed using the 
\texttt{emcee}  software \citep{emcee}, which is a Python code that 
provides a Markov  Chain Monte Carlo (MCMC) implementation of the 
affine-invariant ensemble  sampler of \cite{goodman2010}. The final 
value we obtain from the  analysis of the combined scale is 
$b_2=1.2\pm0.5$. 

\subsection{The advantage of cluster as tracers}
\label{sec:BOSScomparison}
We compare our results both with independent 
estimates and with theoretical forecasts. 
\cite{lazeyras2016} provided a theoretical relation 
between the expected values of $b_1$ and $b_2$ from N-body 
simulations. Using the constraints on $b_1$ we obtained from the 
analysis of the 2PCF and their Eq. 5.2, we obtain $b_2=1.70\pm0.08$, 
consistent with our measurement. 
Comparing our constraints with other analyses of real galaxy catalogs 
in the same redshift range, we find instead some differences, which 
are, however, mostly ascribable to a difference in the tracers 
analyzed. 
\cite{slepian2017}, from the analysis of the 3PCF of SDSS-DR12 CMASS 
galaxies, found $b_2(z=0.565)=0.3\pm0.7$; \cite{marin2013} measured 
the 3PCF of the WiggleZ Dark Energy Survey in three redshift bins, 
finding $b_2(z=0.35)=-0.36^{+0.11}_{-0.08}$, 
$b_2(z=0.55)=-0.41^{+0.09}_{-0.08}$, and 
$b_2(z=0.68)=-0.48^{+0.14}_{-0.12}$. These discrepancies in the
non-linear bias can be explained by taking into account that in 
\cite{slepian2017} the sample consisted in very massive galaxies
(log$(M/M_{\odot}\sim$11.3) at $z\sim0.5$ \citep{maraston2013} 
but still less biased than our tracers, while the WiggleZ survey 
was mostly focused in selecting star-forming galaxies at 
$z\sim0.5$ \citep{marin2013}. The linear bias is $b_1=2.23\pm0.06$, 
$0.72\pm0.14$, $0.99^{+0.10}_{-0.09}$, and $1.06^{+0.16}_{-0.18}$,
respectively, significantly lower than the one of our clusters. 
If we consider the previously discussed $b_2(b_1)$ relation, we find
that this difference in linear bias compensates for the differences in
the non-linear bias results.

It is interesting to notice that while this analysis, given the 
current number of galaxy clusters available, is shot-noise dominated, 
it already shows clearly the advantage of a galaxy cluster sample with
respect to a sample with a smaller bias. On one side, selecting 
objects in the peak of matter over-densities significantly reduces
the effects due to non-linear dynamics, as also found by 
\cite{marulli2020}. On the other side, to assess more quantitatively
the improvement of using such a sample we also perform a complementary
analysis. We use the publicly available SDSS-BOSS galaxy and 
random catalogs\footnote{\url{https://data.sdss.org/sas/dr12/boss/lss/}},
and extract from these a sample with the exact same statistics and
redshift distribution of our cluster sample. In this way, the only
differences that we will find in the results can be attributed to the 
different tracers. We analyze this catalog with the same method
described in Sect.~\ref{sec:measzeta}, and as a first result we 
confirm that this sample is characterized by a smaller bias, 
$b_1=1.95\pm0.02$. We then compare the percentage errors associated
to the measured reduced 3PCF, and find that the cluster catalog
has an error on average $\sim$20\% smaller than the galaxy catalog.
This results in a constraint on $b_2$ from the fit of the reduced
3PCF with errors $\sim$20\% smaller. This highlights the gain in using
more biased tracers in the analysis.

%%%%%%%%%%%%%%%%%%%%%%%%%%%%%%%%%%%%%%%%%%%%%%%%%%%%%%%%%%%%%%%%%%%%%%%%%%%%%%%
%%%%%%%%%%%%%%%%%%%%%%%%%%%%%%%%%%%%%%%%%%%%%%%%%%%%%%%%%%%%%%%%%%%%%%%%%%%%%%%

\section{Detecting the BAO peak in the 3PCF of galaxy clusters}
\label{sec:BAOdet}

\begin{figure}[t!]
  \includegraphics[width=0.48\textwidth]{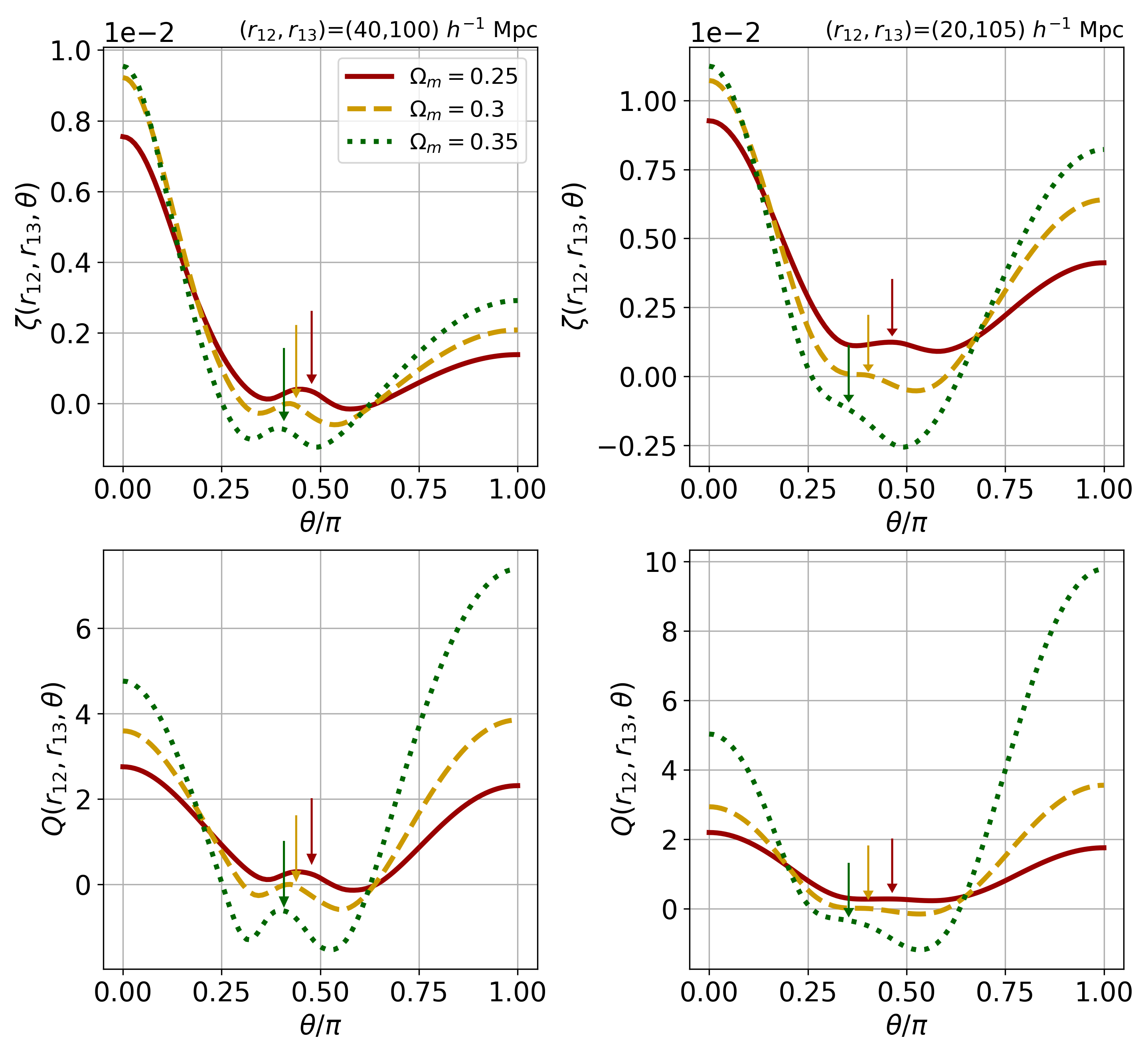}   
\caption{Examples of the BAO detectability in the 3PCF as a function of $\Omega_m$ (represented by different line colors, as shown in the legend) for the connected 3PCF (upper panels) and for the reduced 3PCF (lower panels). Different scale configurations are shown, $(r_{12},r_{13})=$ (40,100) $h^{-1}$ Mpc (left panels) and (20,105) $h^{-1}$ Mpc (right panels). In each panel, a colored arrow show the 
  position of the BAO peak for each model.}
  \label{fig:BAO_det_param}
\end{figure}

The BAO feature appears as a clear peak around 100 $h^{-1}$ Mpc in 
the 2PCF. The peculiar shape of the 3PCF, however, makes a bit more 
difficult to clearly detect this peak, since both $\zeta(\theta)$ and 
$Q(\theta)$ are characterized by a U- or V-shape, with a sharp dip 
around $\theta\sim\pi/2$ (see, e.g., Fig.~\ref{fig:scales_b1b2}). 
Therefore, when probing BAO scales, the shape of the 3PCF will show a 
complex behavior, with different features depending on the relative 
depth and height of the dip and peak.

\begin{figure*}[ht!]
  \includegraphics[width=0.99\textwidth]{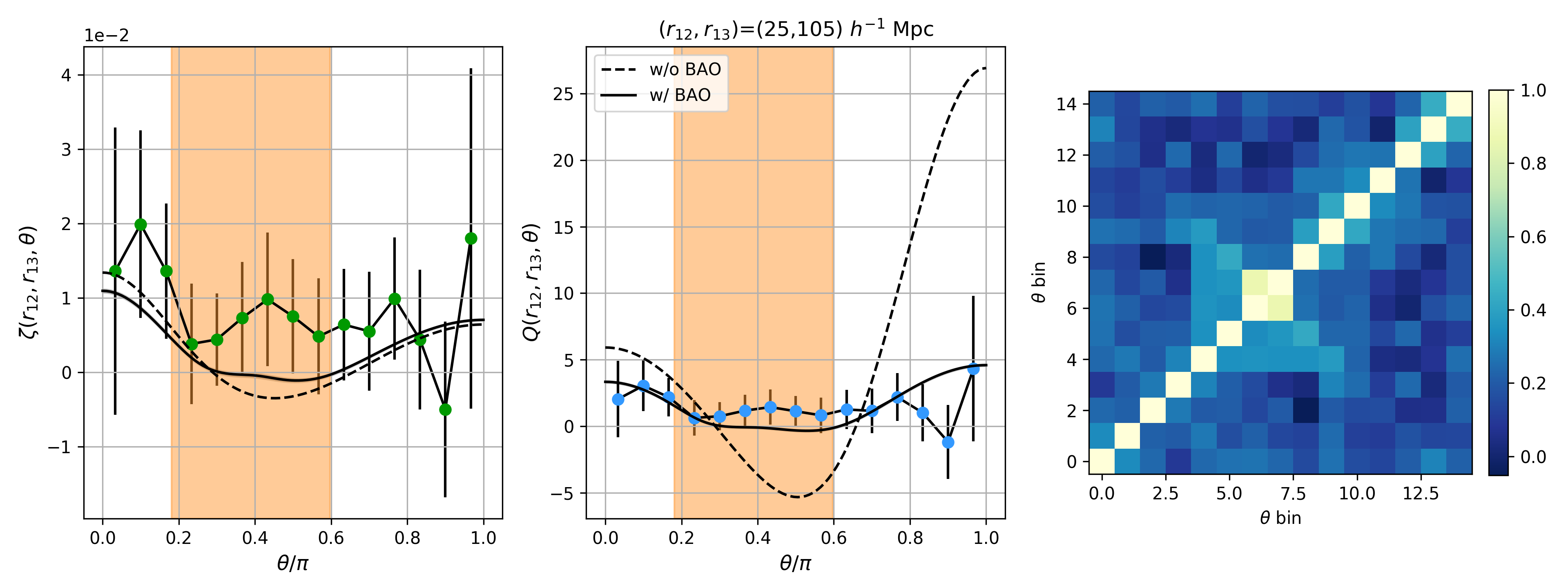}
  \includegraphics[width=0.99\textwidth]{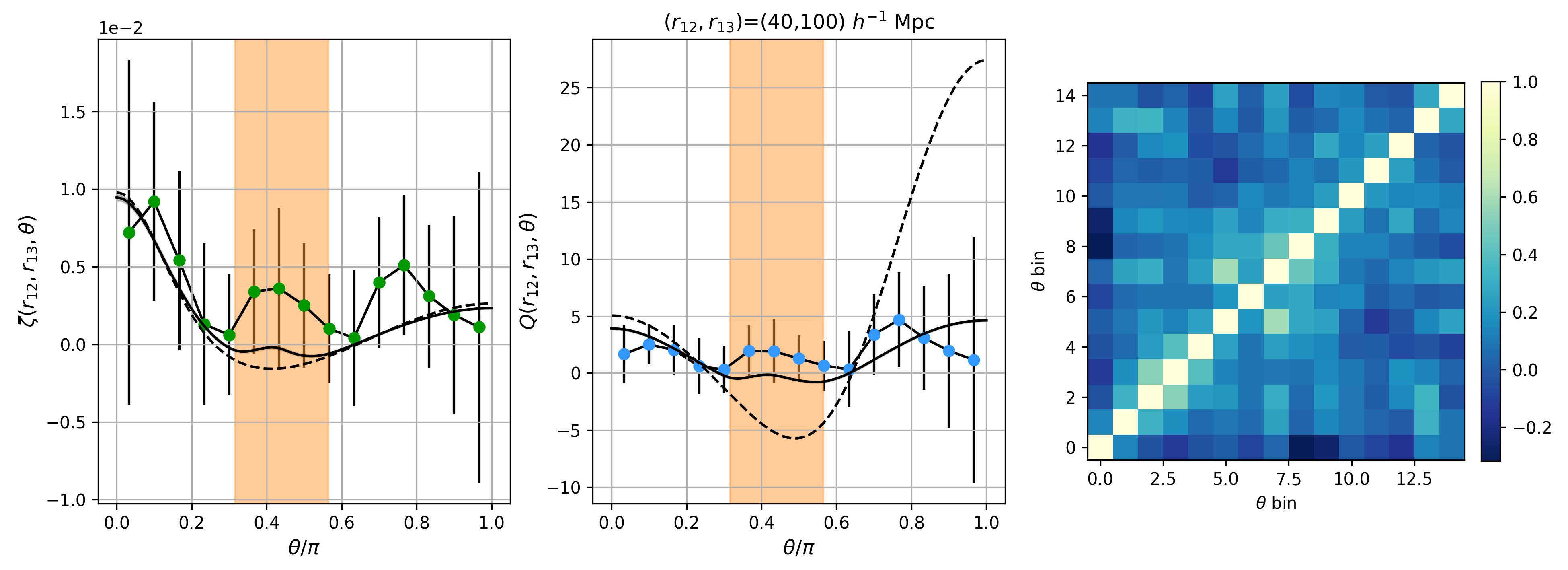}
  \includegraphics[width=0.99\textwidth]{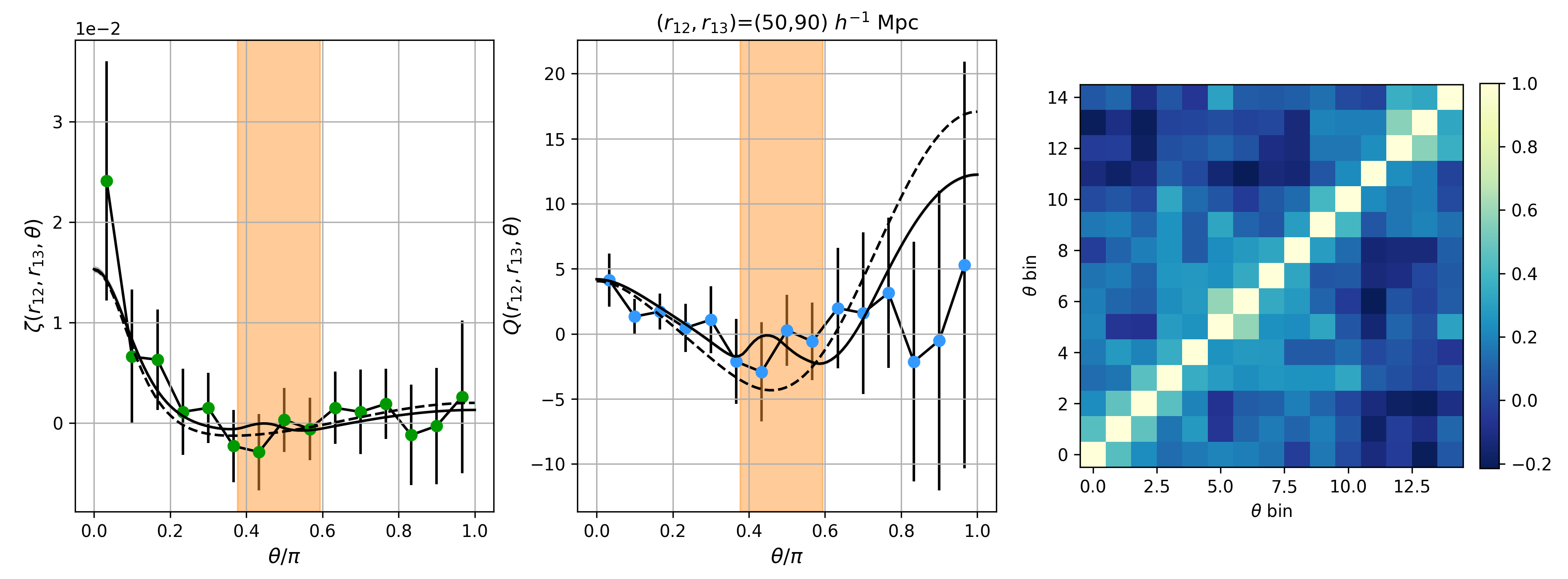}
  \caption{3PCF measurements for SDSS galaxy clusters at BAO
    scales. Left panels: the connected 3PCF; central panels: the
    reduced 3PCF; right panels: the correlation matrices of the
    reduced 3PCF. The configurations chosen are, from top to bottom,
    $(r_{12},r_{13})=$ (20, 105), (40,100), and (50, 90) $h^{-1}$ Mpc,
    respectively. The orange shaded area shows the region $85\leq
    r_{23}$[$h^{-1}$ Mpc]$\leq115$, to highlight the expected position
    of the BAO peak. The black lines show the model at the best-fit
    parameters obtained by fitting 3PCF as smaller scales
    (Sect.~\ref{sec:measbias}), while the dashed line is a model
    assuming the same parameters, but without the BAO peak. The BAO 
    peak can be detected in the reduced 3PCF with a clearer evidence 
    from top to bottom panels as the difference between the two models.}
    \label{fig:BAO}
\end{figure*}

In Fig.~\ref{fig:BAO_det_param}, we show an example of how the results
may appear in different configurations, for both the connected and 
reduced 3PCFs, for different values of $\Omega_m$ (keeping fixed all 
other cosmological parameters, including the power spectrum 
normalization). We used \texttt{CBL} to produce 3PCF models, focusing 
in particular on the two configurations$(r_{12},r_{13})=$ (40,100) and 
(20,105) $h^{-1}$ Mpc, representing two typical behaviors of the 3PCF.  
We find that, at fixed configuration, the BAO peak is sharper and more 
detectable in the reduced 3PCF rather than in the connected 3PCF. 
Cosmological parameters have also an impact on the BAO visibility, with 
a higher peak corresponding to a higher value of $\Omega_m$. As for the
dependence on the configuration, we find that the dip of the 3PCF and
the BAO peak can combine themselves in two ways: they can be of the
same order of magnitude, thus canceling out each other, and giving a
flat 3PCF shape for $\theta\sim\pi/2$ (as can be seen e.g. in the case
$(r_{12},r_{13})=$ (40,100) $h^{-1}$ Mpc), or the BAO peak can
dominate, resulting in a small peak embedded in the dip (as e.g. for
the case $(r_{12},r_{13})=$ (20,105) $h^{-1}$ Mpc). This effect is due
to the fact that it is possible to choose configurations that
concentrate the BAO signal on a smaller or on a wider $\theta$ range:
in the case of $(r_{12},r_{13})=$ (20,105) $h^{-1}$ Mpc, the third
side, $r_{23}$, ranges between 85 and 125 $h^{-1}$ Mpc, and the BAO
peak is spread approximately on the entire $\theta$ range, while for
$(r_{12},r_{13})=$ (40,100) $h^{-1}$ Mpc the BAO signal is
concentrated in $0.3\leq\theta\leq0.5$, with higher visibility.

For the above reasons, we decided to focus on the three configurations
discussed in Sect.~\ref{sec:measzeta}, in order to exploit different
possibilities to detect the BAO peak. The results are shown in
Fig.~\ref{fig:BAO} for configurations progressively concentrating the
BAO signal on a smaller range of angles. First, we recover that the
less concentrated the BAO peak is, the flatter the resulting 3PCF
spread over a smaller number of bins is that the final signal-to-noise
ratio of the measurement is smaller, and the evidence of a peak less
clear. We compare the measurements of the 3PCF with theoretical models
assuming the best-fit parameters estimated in the previous analysis of
Sect.~\ref{sec:measbias} (solid lines). Here, we remind that we
decided to fit $Q(\theta)$ rather than $\zeta(\theta)$, therefore some
discrepancies between models and data in the connected 3PCF are
expected. We find that the obtained models well reproduce the measured
reduced 3PCF, with a reduced $\chi^2$ between 0.3 and 0.7; the small
values of the reduced $\chi^2$ are caused by the large errors
associated to the 3PCF especially at the largest angles, due to the
smaller number of triplets at the largest probed scales.

Figure \ref{fig:BAO} presents also the models obtained with the same
parameters, but from a no wiggle power spectrum that does not include 
the BAO peak obtained with the CBL from the prescriptions of 
\cite{eisenstein1998}. We notice that while
in the connected 3PCF the differences between the two models are only
at BAO scales, as expected, in the reduced 3PCF a significant
deviation appears also for $\theta>0.7$. This effect is due to the
fact that $Q(\theta)$ is defined as the connected 3PCF normalized by a
combination of $\xi^2(r)$ (see Eq.~\eqref{eq:3PCF}). At the large
scales where this deviation is detected ($r>110$ $h^{-1}$ Mpc), the
2PCF becomes smaller and smaller (see Fig.~\ref{fig:2pcf}), and
eventually close to zero due to random fluctuations caused by Poisson
noise. This produces the large variations that are observed in
$Q(\theta)$, but not in $\zeta(\theta)$.

To quantify the detectability of the BAO peak, we use the same
approach adopted in \cite{slepian2017}, estimating the difference in
$\chi^2$ between the best-fit models including or not the BAO,
$\rm\Delta \chi^2_{noBAO-BAO}$. To avoid being biased by the
previously discussed fluctuations in the 2PCF, we restrict our
$\chi^2$ estimate to the angles corresponding to $85\leq r$[$h^{-1}$
  Mpc]$\leq105$ (as also highlighted by the orange shaded area in
Fig.~\ref{fig:BAO}), that is close to the position of the BAO
peak. The results are reported in Tab.~\ref{tab:chi2}.

We find that for all the configurations probed, the model including
the BAO reproduces the observed reduced 3PCF better than the one
without the BAO, with a difference $\rm\Delta \chi^2_{noBAO-BAO}$
between $\sim$2 and $\sim$75. The configuration that shows the largest
evidence for the presence of the BAO peak is $(r_{12},r_{13})$ =
(25,105) $h^{-1}$ Mpc, where the BAO signal is spread over a larger
number of bins. In the other configurations, the BAO feature is
squeezed in a smaller number of bins and, as expected from previous
considerations, it shows some hints of a peak both in the connected
and in the reduced 3PCF. These configurations, however, also show a
larger error, and therefore the resulting evidence is lower, but still
preferring the presence of a BAO peak. 

These data represent the first detection of the BAO signal in the 3PCF
of galaxy clusters. Moreover, they also pave the way to select the best 
configurations to detect the BAO signal in the 3PCF, showing that it 
has a stronger signal in the reduced than in the connected 3PCF, and 
that the configurations maximizing the signals are the ones for which 
the BAO peak is sampled on a larger number of bins. To further 
systematize these results, in the next section we will discuss a 
theoretical framework to estimate the SNR of the BAO peak in the 3PCF 
as a function of the configuration probed.

\begin{table}[t!]
\centering
\caption{Differences in $\chi^2$ between models with and without BAO
  (assuming Planck18 cosmology). The $\Delta \chi^2$ have been
  estimated in the range of separations interested by the BAO feature, 
  corresponding to the orange band in Fig. \ref{fig:BAO}.}
\label{tab:tab2}
\begin{tabular}{lc}
\hline
scales & $\rm\Delta \chi^2_{noBAO-BAO}$\\
\hline
$r_{12}=25$ $h^{-1}$ Mpc, $r_{13}=105$ $h^{-1}$ Mpc & 74.9\\
$r_{12}=40$ $h^{-1}$ Mpc, $r_{13}=100$ $h^{-1}$ Mpc & 21.6\\
$r_{12}=50$ $h^{-1}$ Mpc, $r_{13}=90$ $h^{-1}$ Mpc & 2.1\\
\hline
\end{tabular}
\label{tab:chi2}
\end{table}

%%%%%%%%%%%%%%%%%%%%%%%%%%%%%%%%%%%%%%%%%%%%%%%%%%%%%%%%%%%%%%%%%%%%%%%%%%%%%%%
%%%%%%%%%%%%%%%%%%%%%%%%%%%%%%%%%%%%%%%%%%%%%%%%%%%%%%%%%%%%%%%%%%%%%%%%%%%%%%%
\begin{figure*}[t!]
  \includegraphics[width=0.95\textwidth]{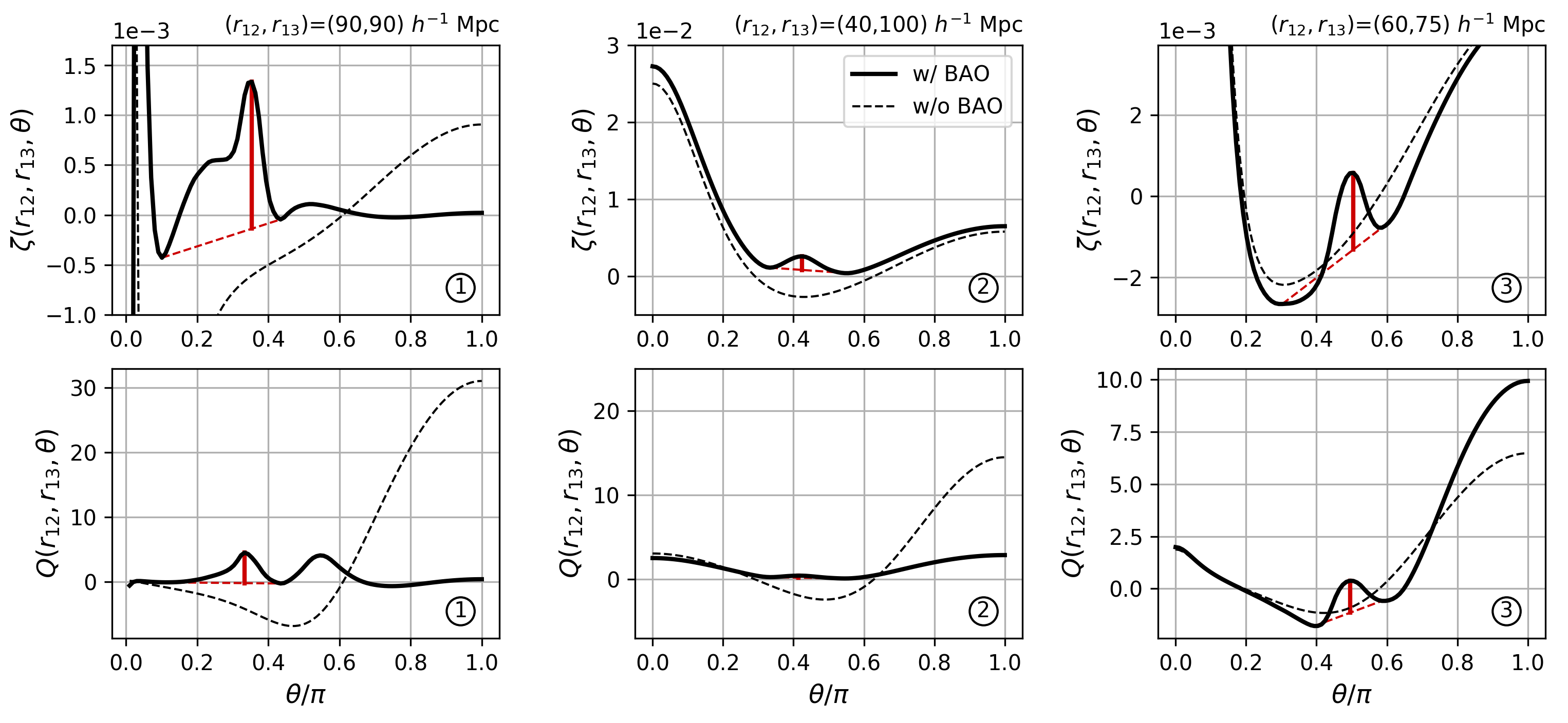}
  \caption{Examples of theoretical 3PCF estimates. The top panels show
    the connected 3PCF at three scales, namely $(r_{12},    
    r_{13})$=(90,90), (40,100), and (60,75) $h^{-1}$ Mpc (panels from 
    left to right), while the lower panels show the associated reduced 
    3PCFs. In all panels, black solid lines show the model including the 
    BAO, while the dashed lines the model without the BAO. The red lines 
    show the estimated {\it BAO contrast}, as discussed in    
    Sect.~\ref{sec:BAO_mod_sign}. Each panel is also labelled with a    
    number, showing the position of the configuration in the following   
    Figs.~\ref{fig:BAO_det_theory} and \ref{fig:BAO_detpeak_theory}.}
  \label{fig:3PCF_mod_examples}
\end{figure*}

%%%%%%%%%%%%%%%%%%%%%%%%%%%%%%%%%%%%%%%%%%%%%%%%%%%%%%%%%%%%%%%%%%%%%%%%%%%%%%%
%%%%%%%%%%%%%%%%%%%%%%%%%%%%%%%%%%%%%%%%%%%%%%%%%%%%%%%%%%%%%%%%%%%%%%%%%%%%%%%

\section{BAO detectability in the 3PCF of Euclid clusters}
\label{sec:BAO}
As previously discussed, one of the main results of this work is that
the BAO peak in the 3PCF can be more or less evident depending on the
considered configuration. For this reason, in Sect.~\ref{sec:analysis}
we analyzed three different configurations of scales. In this Section,
we provide a framework in order to forecast which are the most
favorable configurations to detect the BAO peak in the 3PCF, both for
the connected and the reduced 3PCFs, and apply it to the case of the 
Euclid mission as an example of a future wide survey of galaxy clusters.

For this purpose, we need three ingredients: (i) a theoretical
estimate of the BAO signal in the 3PCF at various scales, (ii) a
theoretical estimate of the expected corresponding error, and (iii) an
estimate of the associated SNR.

\subsection{The BAO signal in the 3PCF}
\label{sec:BAO_mod_sign}
As a first step, we used the \texttt{CBL} to compute theoretical 3PCF
models following \cite{barriga2002}. 
To provide forecasts on the detectability of the BAO peak with Euclid
clusters, we base our analysis on the expected number density and mass
for these objects estimated in the work of \cite{sartoris2016}. 
We assume a Euclid-like survey that will detect approximatively 
2$\times10^5$ clusters with S/N>3 in the Euclid photometric survey 
\citep{euclid} in the range $0.2<z<2$, with masses 
log$(M/M_{\odot}\gtrsim$13.9 and a mean redshift $<z>\sim1$; the 
effective bias $b_1$ has been estimated from the scale relation by 
\cite{tinker2010} ($b_1=5.4$), and the non-linear bias parameter $b_2$ 
from the relation by \cite{lazeyras2016} ($b_2=17.4$), assuming a 
Planck18 cosmology.

The models have been created considering every possible closed triangle
configurations with $r_{12}$ and $r_{13}$ between 30 and 115 $h^{-1}$
Mpc, with a binning of 5 $h^{-1}$ Mpc. The third side $r_{23}$ spans
all the possible allowed values between $r_{min}=r_{13}-r_{12}$ and
$r_{max}=r_{13}+r_{12}$; therefore, by definition for some
combinations the BAO peak will not be visible in $\zeta(\theta)$ and
$Q(\theta)$, but in the majority it will\footnote{For example, the
  combination $(r_{12}, r_{13})$=(30,30) $h^{-1}$ Mpc will not allow
  to see the BAO peak, since both in $\zeta(r_{12}, r_{13}, \theta)$
  and $Q(r_{12}, r_{13}, \theta)$ the various angles will map to a
  third side in the range $0<r_{23}$[$h^{-1}$ Mpc]$<60$.}.

To measure the BAO signal in the 3PCF, two approaches have been
considered.
\begin{itemize}
\item {\bf BAO from model comparison.} The first approach, similar to
  the one considered in Sect.~\ref{sec:BAOdet}, is to measure the
  difference between 3PCF models with and without the inclusion of the
  BAO signal. In this case, we average the values of the 3PCF in the
  range of $\theta$ such as $85<r_{23}$[$h^{-1}$ Mpc]$<115$, i.e. the
  scales for which the BAO peak is observed at these redshifts, for
  both models, and estimate the difference between these two
  values. We define this value as {\it BAO difference}. This has been
  done for both the connected and the reduced 3PCFs.
\item {\bf BAO contrast.} In the second approach, to be less model
  dependent, we only consider the model including BAO, and measure if
  it is possible to detect the presence of a peak in the 3PCF. As
  previously discussed, this is not a trivial task, since given the
  shape of the 3PCF it means to detect a secondary peak between two
  larger peaks in the 3PCF that usually are at $\theta\sim0$ and
  $\theta\sim\pi$. To do this, we use the modules \texttt{argrelmin}
  and \texttt{argrelmax} of the \texttt{scipy.signal} Python library
  \citep{scipy}, specifically designed to detect relative maxima and
  minima in a function. To define a peak, and, most importantly, its
  height, we therefore use these functions to detect a relative
  maximum between $0<\theta<\pi$, and its adjacent relative
  minima. These relative minima are used to define a baseline for the
  peak by linearly interpolating between them, and the height of the
  peak is then measured as the distance between the relative maximum
  and the baseline. We define this value as the {\it BAO contrast}. As
  for the previous approach, this has been estimated both on
  $\zeta(\theta)$ and $Q(\theta)$.
\end{itemize}
For illustrative purposes, in Fig.~\ref{fig:3PCF_mod_examples} we show
some examples of the generated 3PCFs, including both the models with
and without the BAO signal, and a visual representation of the {\it
  BAO contrast}.

\subsection{A theoretical estimate of the 3PCF covariance}
\label{sec:BAO_mod_err}
To estimate the errors, we consider the theoretical covariance as
defined by \cite{slepian2015} (see their Sect.~6.2, and Eq.~51),
and included in the \texttt{CBL}. 
This formula provides an analytical estimate of the covariance depending
on two main parameters: the number of objects and the volume of the 
survey. We consider the volume that will be covered by the Euclid mission
with its 15,000 square degrees between $0.2<z<2$, and 2$\times10^5$
clusters as obtained from the forecasts of \cite{sartoris2016}.

The covariance matrices have been calculated for all the simulated configurations (as discussed in Sect.~\ref{sec:BAO_mod_sign}), and they have been used to estimate the expected errors at the BAO scale, by averaging the errors in the same range used to estimate the BAO signal.

\begin{figure}
  \includegraphics[width=0.48\textwidth]{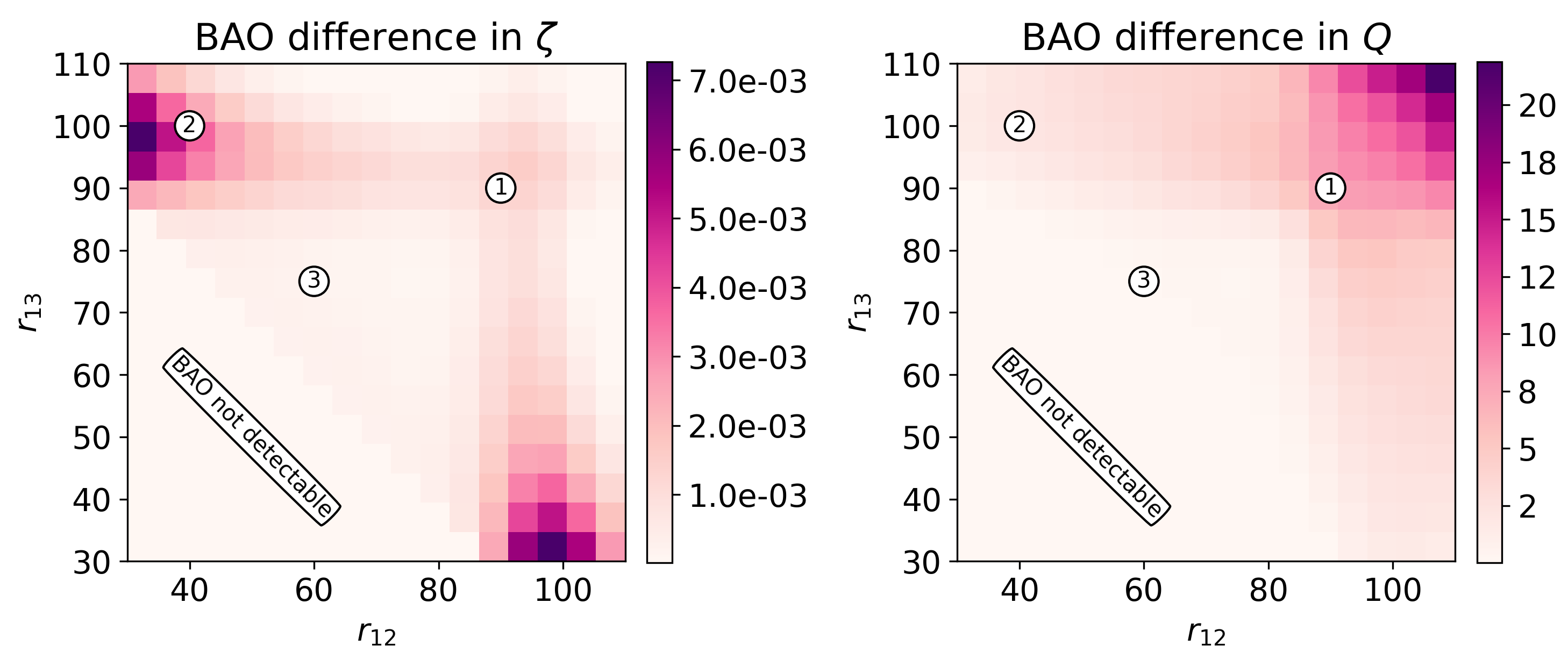}
  \includegraphics[width=0.48\textwidth]{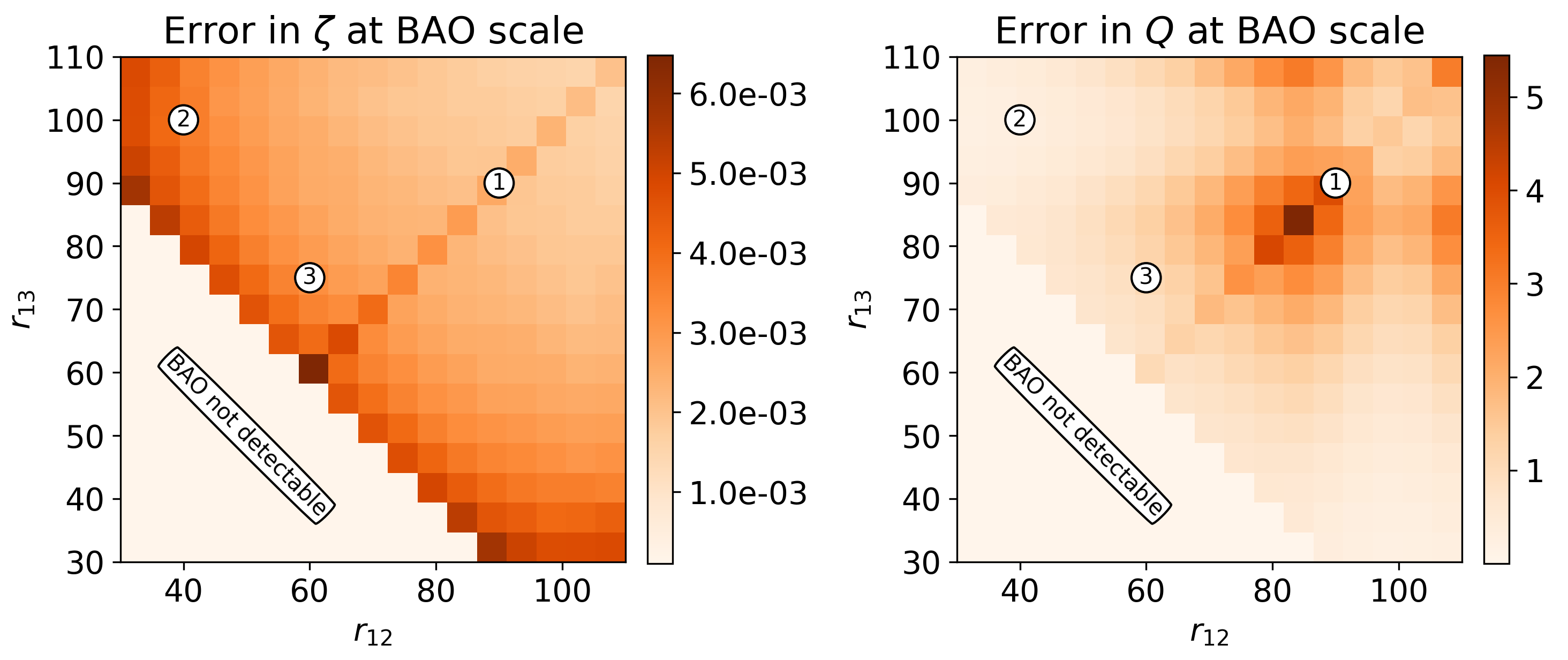}
  \includegraphics[width=0.48\textwidth]{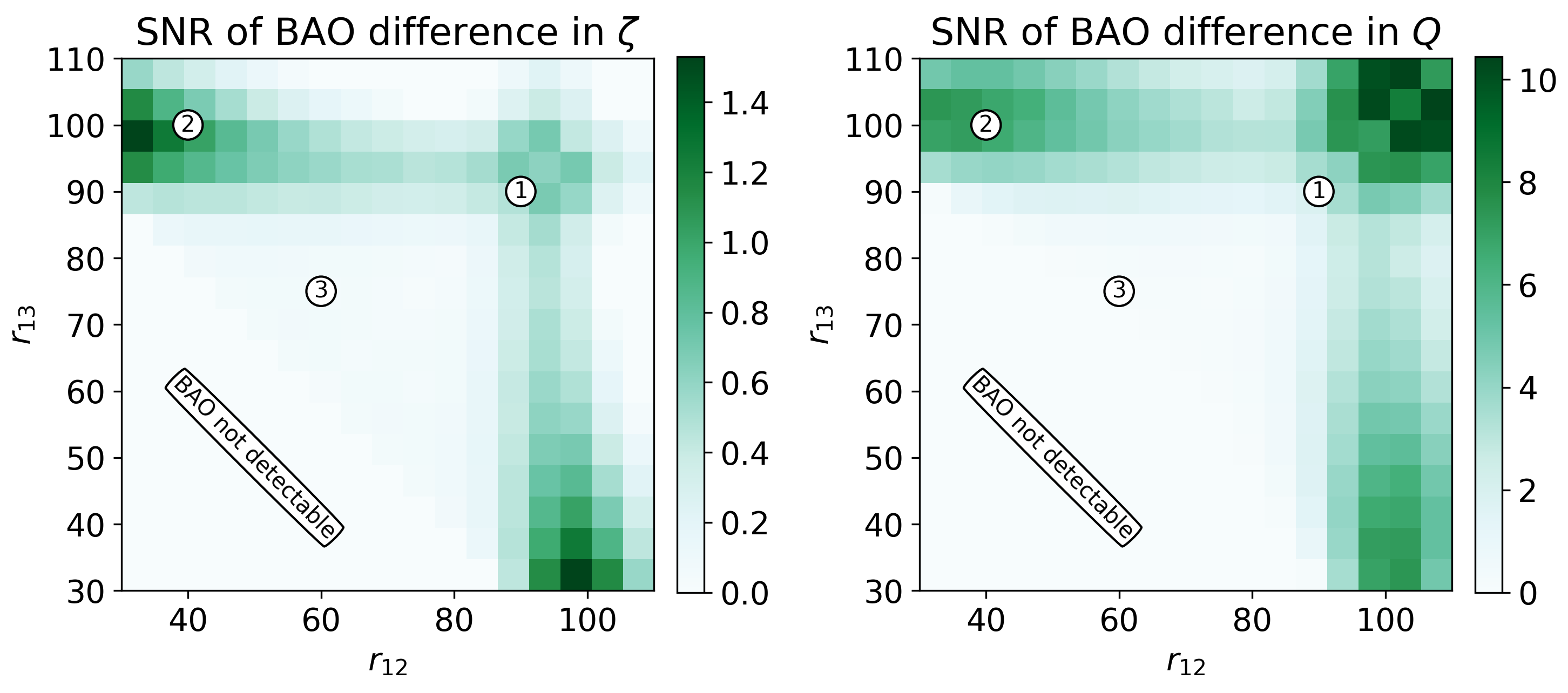}
   
  \caption{BAO detectability in the 3PCF for the {\it BAO difference}
    case for both the connected and reduced 3PCFs (left and right 
    panels, respectively). From top to bottom, the different panels 
    show: the detectability BAO signature, estimated as the 
    difference between models with and without BAO; the expected 
    errors (as described in Sect.~\ref{sec:BAO_mod_err}); the SNR of 
    the BAO peak. The region where the BAO peak is not visible is 
    shown in the bottom-left part of each panel. We indicate with 
    labels the position of the configurations shown in 
    Fig.~\ref{fig:3PCF_mod_examples}.}
    \label{fig:BAO_det_theory}
\end{figure}

\begin{figure}[t!]
  \includegraphics[width=0.48\textwidth]{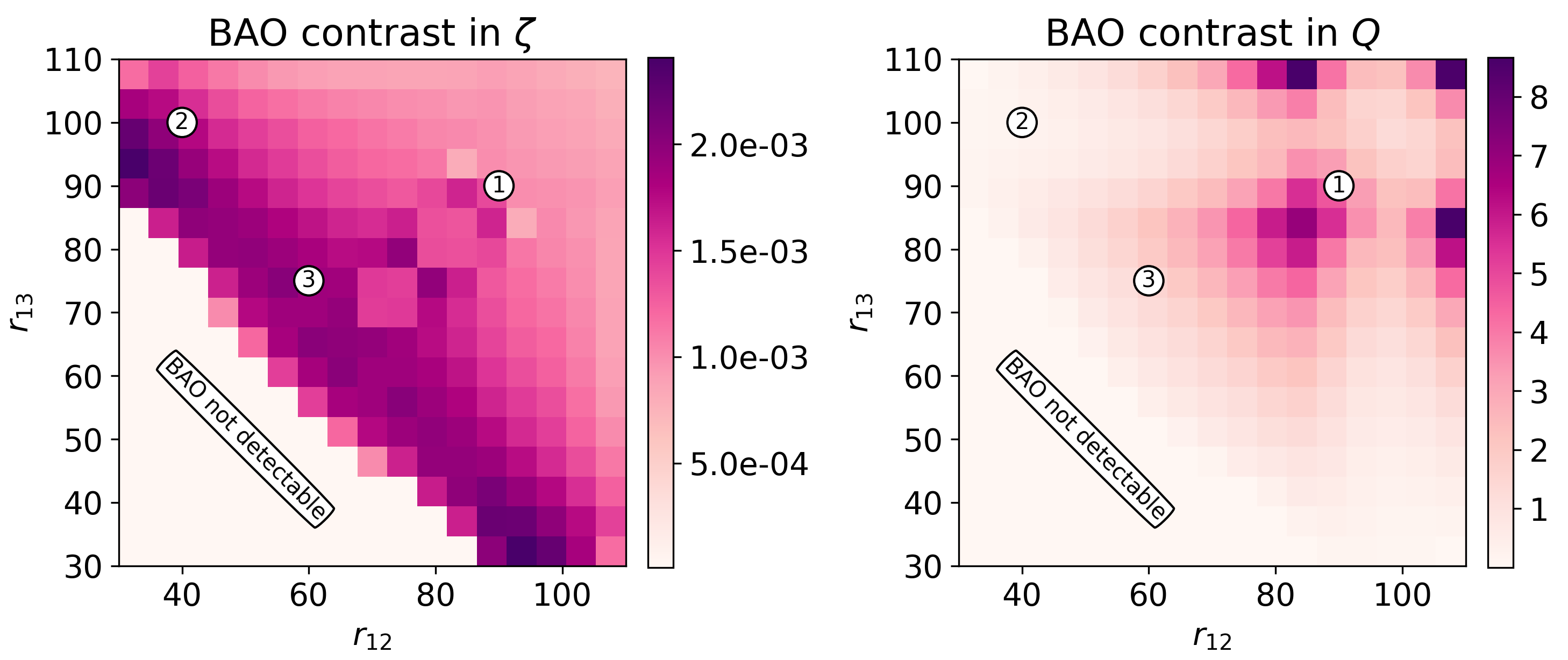}
  \includegraphics[width=0.48\textwidth]{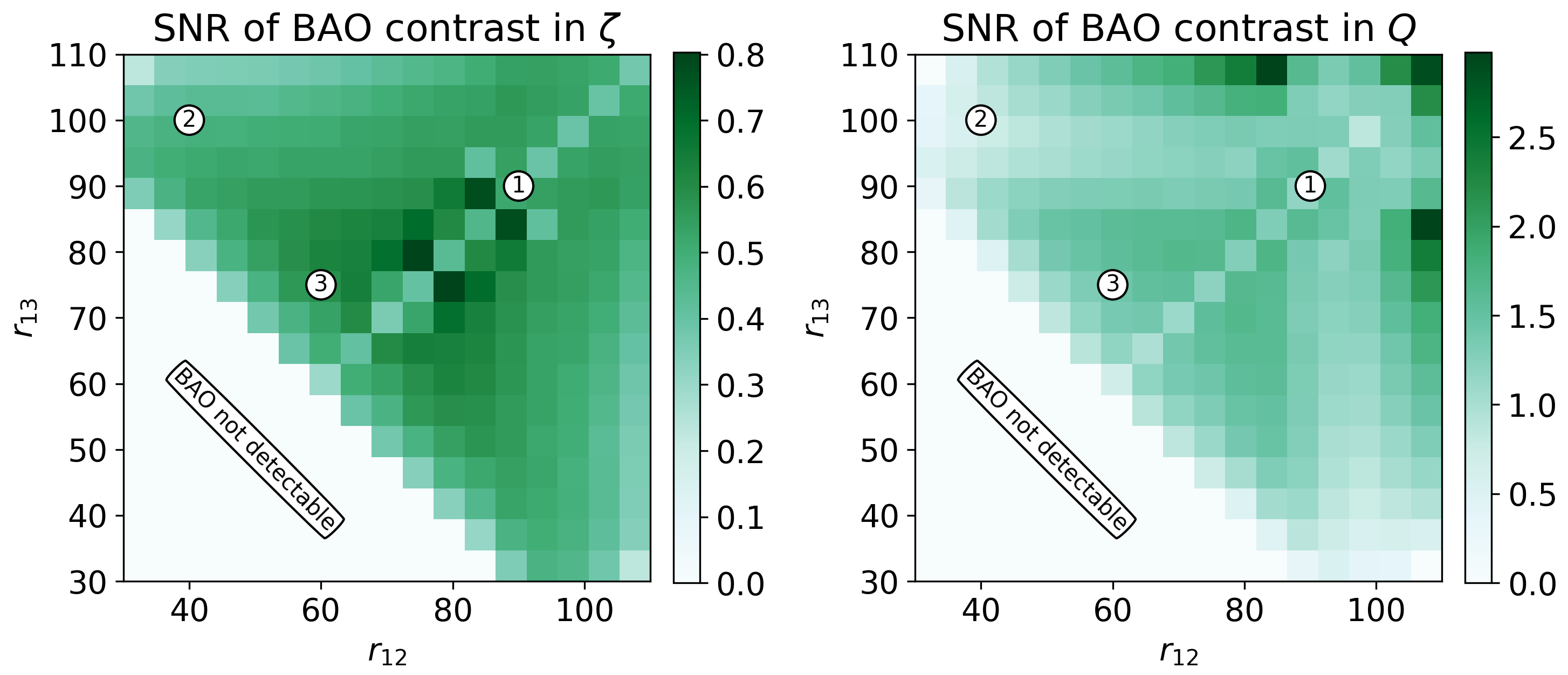}
  \caption{Same as Fig. \ref{fig:BAO_det_theory}, but showing the
    detectability of the BAO peak as its contrast with respect to the
    overall shape of the 3PCF. This measurement would allow a
    determination of BAO position.}
  \label{fig:BAO_detpeak_theory}
\end{figure}

Finally, we estimate the SNR of the BAO signal by dividing the estimated value, either the {\it BAO difference} or the {\it BAO contrast}, and the estimated error. In this way, we obtain the six maps that are shown in Figs.~\ref{fig:BAO_det_theory} and \ref{fig:BAO_detpeak_theory}, one for the signal, one for the error and one for the SNR, for both the connected and the reduced 3PCFs. These maps can be used to forecast which are the preferable configurations that maximize the BAO peak in the 3PCF.

\subsection{Discussion}
Figure \ref{fig:BAO_det_theory} reports the results for the {\it BAO
difference} case. As can be seen, even for the signal itself, the
result is quite different in the connected and reduced 3PCF cases, the
former one being maximal for configurations close to $(r_{12},
r_{13})$=(40,100) $h^{-1}$ Mpc, while the second one being higher for
the equilateral configuration where all three sides are closer to the
BAO scale. The errors, however, follow a different distribution, being
larger for isosceles configurations with a smaller size in
$\zeta(\theta)$, while presenting a peak around $(r_{12},
r_{13})$=(80,80) in $Q(\theta)$.  As a result, the distribution of the
expected SNR is very different between the connected and reduced 3PCF.
In the first case, we find that it is maximum for configurations close
to $(r_{12}, r_{13})$=(40,100) or (90,90) $h^{-1}$ Mpc, but with
smaller values of the SNR, SNR$\sim1.5$ at maximum. On
the other hand, the results obtained for the reduced 3PCF are more
promising, with high SNR for configurations $(r_{12},
r_{13})$=(40,100) and (90,90) $h^{-1}$ Mpc, and values of the order of
SNR$\sim6-10$.  This reflects also the results found from the
analysis of the data shown in Fig.~\ref{fig:BAO}, where it is evident
that the difference between the models with and without the BAO is
very small for $\zeta(\theta)$, but much more significant for
$Q(\theta)$. This effect highlights even more the advantage of using
the reduced 3PCF.

The results obtained from the analysis of the {\it BAO contrast} of
Fig.~\ref{fig:BAO_detpeak_theory} show, on average, a smaller
predicted value of the SNR, with values of the order of 0.8-2.5 for 
$\zeta(\theta)$ and $Q(\theta)$ respectively. This is a
direct consequence of the fact that the measured {\it BAO contrast} is
for all configurations smaller than the {\it BAO difference}. It is
interesting to notice, however, that the configurations where the SNR is
maximized is significantly different in the two approaches. The
configurations explored in our analysis (discussed in
Sect.~\ref{sec:BAO}) have been explicitly chosen to capture this
variety. While having a smaller probability of detection, we think
that the {\it BAO contrast} approach is nevertheless promising 
being less model-dependent, especially for future surveys with which 
the expected error will be significantly smaller due to the much larger 
statistics. 

%%%%%%%%%%%%%%%%%%%%%%%%%%%%%%%%%%%%%%%%%%%%%%%%%%%%%%%%%%%%%%%%%%%%%%%%%%%%%%%
%%%%%%%%%%%%%%%%%%%%%%%%%%%%%%%%%%%%%%%%%%%%%%%%%%%%%%%%%%%%%%%%%%%%%%%%%%%%%%%

\section{Conclusions}
\label{sec:concl}
In this paper, we explored for the first time the possibility 
of detecting the BAO signal in the 3PCF of galaxy clusters.

We analyze the catalog of galaxy clusters obtained cross-correlating
the photometric cluster catalog by \cite{wen2012} with the spectroscopic
information from SDSS-DR12. In this way, we end up obtaining
information on the BCGs of each cluster with an accurate estimate of its 
redshift, fundamental for clustering studies. The final catalog comprises
72,563 objects in the redshift range $0.1<z<0.7$, with a mean redshift
$<z>=0.38$. We measure the 3PCF of this sample to
provide constraints on the properties of this sample in terms of
bias and of the detectability of the BAO peak signal.

Our main results can be summarized as follows:
\begin{itemize}
\item We measured both the connected 3PCF, $\zeta(\theta)$, and the
  reduced 3PCF, $Q(\theta)$, for a variety of configurations, spanning
  from intermediate scales ($r_{23}=20$ $h^{-1}$ Mpc) up to very large
  scales ($r_{23}=80$ $h^{-1}$ Mpc). We also measured for the same
  sample the monopole of the 2PCF $\xi(s)$ in the range
  $10<s$[$h^{-1}$ Mpc]$<80$.
\item We analyzed the 2PCF to get an estimate of the linear bias of
  the sample, obtaining $b_1=2.75\pm0.03$.
\item We used the reduced 3PCF measurements at intermediate scales to
  place constraints on the non-linear bias of the galaxy clusters in the 
  sample. We considered a two-parameter model from Eq.~\eqref{eq:Q_mod} 
  including $b_1$ and $b_2$, since we verified that it was not feasible 
  to set constraints on the additional tidal bias $g_2$. From the 3PCF
  modeling, combined with the information on the linear bias from the
  2PCF, we obtained a constraint on the non-linear bias
  $b_2=1.2\pm0.5$, considering a joint fit to the scales
  $20<r_{23}$[$h^{-1}$ Mpc]$<80$.
\item Considering the obtained best-fit values $b_1$ and $b_2$, we
  analyzed the reduced 3PCF at large scales, exploring different
  configurations for which the BAO peak is present in $Q(\theta)$. We
  compared theoretical models of the 3PCF including or not the BAO
  feature, finding a significant detection of the BAO peak, with the
  models considering BAO always having a better $\chi^2$, with a
  difference $\Delta \chi^2$ between 2 and 75, depending on the
  configuration.
\item To demonstrate the advantage of using galaxy clusters as tracers,
  we used publicly available SDSS-BOSS galaxies and random catalogs to 
  select a catalog with the same properties in terms of numbers ad 
  redshift distribution of the cluster one, and measured the 3PCF on 
  this catalog. We found that on average the percentage errors on $Q$
  is 20\% smaller for the cluster catalog, having, as a consequence,
  20\% better constraints on bias parameters. We discussed how this 
  effect is a combination of the higher bias and of the smaller impact of non-linear dynamics on the cluster sample.
\item We developed a framework to theoretically calculate the 
  expected SNR of the BAO feature in the 3PCF of a generic population, 
  and applied it to provide a forecast for clusters detected with the 
  future Euclid mission. We introduced two
  possible parameterizations to define the presence of the BAO peak,
  based (i) on the comparison between models with and without the BAO,
  and (ii) on the direct detection of a peak in the 3PCF. We
  calibrated a theoretical covariance on the expected performance
  to detect clusters that have been forecasted for the Euclid survey,
  and estimated the expected SNR for both the connected and the reduced
  3PCF for a variety of configurations, to find those maximizing its
  value.
\item From this framework, we find that the expected SNR of the BAO
  feature is systematically higher in the reduced 3PCF, $Q(\theta)$,
  than in the connected 3PCF, $\zeta(\theta)$. We also provide the
  configurations that optimize the detection of the BAO signal for 
  a Euclid-like cluster sample.
\end{itemize}

This paper, the second of the $C^3$ project, demonstrates that galaxy 
clusters are very interesting tracers to be analyzed also with higher-
order correlation functions, providing the first significant detection 
of the BAO peak in the 3PCF of galaxy clusters and indicating that this
population can provide a cleaner and stronger signal than normal 
galaxies (under the same conditions of number of objects and 
distribution). In Paper III \cite{veropalumbo2020} we will combine all 
the information that we got from 2PCF and 3PCF exploited in previous 
papers, and show how the combination of the two can provide 
tighter constraints on cosmological parameters, breaking degeneracies 
between them and showing the potential of the 3PCF probe as a 
complementary analysis to measure the BAO signal with respect to the 
more widely used 2-point statistics.

While the analysis is currently shot-noise dominated due to the
available statistics, future surveys will significantly increase
current-day statistics, also expanding the redshift coverage. As a few
examples, the Dark Energy Survey (DES) has already collected
$\sim7\times10^3$ galaxy clusters \citep{abbott2020} and is expected
to provide $\sim1.7\times10^{5}$ up to $z\sim1.5$ with the final
release; Euclid is expected to detect $\sim2\times10^{5}$ galaxy
clusters across the entire redshift range, of which
$\sim4\times10^{4}$ at $z>1$ \citep{sartoris2016}; and eROSITA
\citep{pillepich2012} will detect $\sim9.3\times10^{4}$ clusters at
$z<1$.

Finally, the framework introduced in Sect.~\ref{sec:BAO} can be useful 
not only to provide forecasts on the expected SNR of the BAO in the 
3PCF for incoming future surveys like Euclid \citep{euclid}, the Vera C. 
Rubin Observatory LSST \citep{lsst2019}, and the Nancy Grace Roman Space 
Telescope \citep{spergel2015}, but also to significantly focus the 
efforts in the search of BAO signal in the 3PCF, at the same time 
maximizing the scientific return while helping in minimizing the 
computational costs.

%%%%%%%%%%%%%%%%%%%%%%%%%%%%%%%%%%%%%%%%%%%%%%%%%%%%%%%%%%%%%%%%%%%%%%%%%%%%%%%
%%%%%%%%%%%%%%%%%%%%%%%%%%%%%%%%%%%%%%%%%%%%%%%%%%%%%%%%%%%%%%%%%%%%%%%%%%%%%%%

\section*{Acknowledgments}
We acknowledge the grants ASI n.I/023/12/0 and ASI n.2018-
23-HH.0. LM acknowledges support from the grant PRIN-MIUR 2017 WSCC32.

{\em Software}: \texttt{CosmoBolognaLib} \citep{CBL};
\texttt{CAMB} \citep{lewis2000}; \texttt{emcee} \citep{emcee};
\texttt{Matplotlib} \citep{hunter2007}; \texttt{scipy} \citep{scipy}.

%%%%%%%%%%%%%%%%%%%%%%%%%%%%%%%%%%%%%%%%%%%%%%%%%%%%%%%%%%%%%%%%%%%%%%%%%%%%%%%

\bibliography{bib}{}

\begin{thebibliography}{}
\expandafter\ifx\csname natexlab\endcsname\relax\def\natexlab#1{#1}\fi
\providecommand{\url}[1]{\href{#1}{#1}}
\providecommand{\dodoi}[1]{doi:~\href{http://doi.org/#1}{\nolinkurl{#1}}}
\providecommand{\doeprint}[1]{\href{http://ascl.net/#1}{\nolinkurl{http://ascl.net/#1}}}
\providecommand{\doarXiv}[1]{\href{https://arxiv.org/abs/#1}{\nolinkurl{https://arxiv.org/abs/#1}}}

\bibitem[{{Abbott} {et~al.}(2020){Abbott}, {Aguena}, {Alarcon}, {Allam},
  {Allen}, {Annis}, {Avila}, {Bacon}, {Bechtol}, {Bermeo}, {Bernstein},
  {Bertin}, {Bhargava}, {Bocquet}, {Brooks}, {Brout}, {Buckley-Geer}, {Burke},
  {Carnero Rosell}, {Carrasco Kind}, {Carretero}, {Castander}, {Cawthon},
  {Chang}, {Chen}, {Choi}, {Costanzi}, {Crocce}, {da Costa}, {Davis}, {De
  Vicente}, {DeRose}, {Desai}, {Diehl}, {Dietrich}, {Dodelson}, {Doel},
  {Drlica-Wagner}, {Eckert}, {Eifler}, {Elvin-Poole}, {Estrada}, {Everett},
  {Evrard}, {Farahi}, {Ferrero}, {Flaugher}, {Fosalba}, {Frieman},
  {Garc{\'\i}a-Bellido}, {Gatti}, {Gaztanaga}, {Gerdes}, {Giannantonio},
  {Giles}, {Grandis}, {Gruen}, {Gruendl}, {Gschwend}, {Gutierrez}, {Hartley},
  {Hinton}, {Hollowood}, {Honscheid}, {Hoyle}, {Huterer}, {James}, {Jarvis},
  {Jeltema}, {Johnson}, {Johnson}, {Kent}, {Krause}, {Kron}, {Kuehn},
  {Kuropatkin}, {Lahav}, {Li}, {Lidman}, {Lima}, {Lin}, {MacCrann}, {Maia},
  {Mantz}, {Marshall}, {Martini}, {Mayers}, {Melchior}, {Mena-Fern{\'a}ndez},
  {Menanteau}, {Miquel}, {Mohr}, {Nichol}, {Nord}, {Ogand o}, {Palmese},
  {Paz-Chinch{\'o}n}, {Plazas}, {Prat}, {Rau}, {Romer}, {Roodman}, {Rooney},
  {Rozo}, {Rykoff}, {Sako}, {Samuroff}, {S{\'a}nchez}, {Sanchez}, {Saro},
  {Scarpine}, {Schubnell}, {Scolnic}, {Serrano}, {Sevilla-Noarbe}, {Sheldon},
  {Smith}, {Smith}, {Suchyta}, {Swanson}, {Tarle}, {Thomas}, {To}, {Troxel},
  {Tucker}, {Varga}, {von der Linden}, {Walker}, {Wechsler}, {Weller},
  {Wilkinson}, {Wu}, {Yanny}, {Zhang}, {Zhang}, {Zuntz}, \& {DES
  Collaboration}}]{abbott2020}
{Abbott}, T.~M.~C., {Aguena}, M., {Alarcon}, A., {et~al.} 2020, \prd, 102,
  023509, \dodoi{10.1103/PhysRevD.102.023509}

\bibitem[{{Alam} {et~al.}(2015){Alam}, {Albareti}, {Allende Prieto}, {Anders},
  {Anderson}, {Anderton}, {Andrews}, {Armengaud}, {Aubourg}, {Bailey}, \&
  et~al.}]{alam2015}
{Alam}, S., {Albareti}, F.~D., {Allende Prieto}, C., {et~al.} 2015, \apjs, 219,
  12, \dodoi{10.1088/0067-0049/219/1/12}

\bibitem[{{Alam} {et~al.}(2017){Alam}, {Ata}, {Bailey}, {Beutler}, {Bizyaev},
  {Blazek}, {Bolton}, {Brownstein}, {Burden}, {Chuang}, {Comparat}, {Cuesta},
  {Dawson}, {Eisenstein}, {Escoffier}, {Gil-Mar{\'\i}n}, {Grieb}, {Hand}, {Ho},
  {Kinemuchi}, {Kirkby}, {Kitaura}, {Malanushenko}, {Malanushenko}, {Maraston},
  {McBride}, {Nichol}, {Olmstead}, {Oravetz}, {Padmanabhan},
  {Palanque-Delabrouille}, {Pan}, {Pellejero-Ibanez}, {Percival}, {Petitjean},
  {Prada}, {Price-Whelan}, {Reid}, {Rodr{\'\i}guez-Torres}, {Roe}, {Ross},
  {Ross}, {Rossi}, {Rubi{\~n}o-Mart{\'\i}n}, {Saito}, {Salazar-Albornoz},
  {Samushia}, {S{\'a}nchez}, {Satpathy}, {Schlegel}, {Schneider},
  {Sc{\'o}ccola}, {Seo}, {Sheldon}, {Simmons}, {Slosar}, {Strauss}, {Swanson},
  {Thomas}, {Tinker}, {Tojeiro}, {Maga{\~n}a}, {Vazquez}, {Verde}, {Wake},
  {Wang}, {Weinberg}, {White}, {Wood-Vasey}, {Y{\`e}che}, {Zehavi}, {Zhai}, \&
  {Zhao}}]{alam2017}
{Alam}, S., {Ata}, M., {Bailey}, S., {et~al.} 2017, \mnras, 470, 2617,
  \dodoi{10.1093/mnras/stx721}

\bibitem[{{Allen} {et~al.}(2011){Allen}, {Evrard}, \& {Mantz}}]{allen2011}
{Allen}, S.~W., {Evrard}, A.~E., \& {Mantz}, A.~B. 2011, \araa, 49, 409,
  \dodoi{10.1146/annurev-astro-081710-102514}

\bibitem[{{Anderson} {et~al.}(2014){Anderson}, {Aubourg}, {Bailey}, {Beutler},
  {Bhardwaj}, {Blanton}, {Bolton}, {Brinkmann}, {Brownstein}, {Burden},
  {Chuang}, {Cuesta}, {Dawson}, {Eisenstein}, {Escoffier}, {Gunn}, {Guo}, {Ho},
  {Honscheid}, {Howlett}, {Kirkby}, {Lupton}, {Manera}, {Maraston}, {McBride},
  {Mena}, {Montesano}, {Nichol}, {Nuza}, {Olmstead}, {Padmanabhan},
  {Palanque-Delabrouille}, {Parejko}, {Percival}, {Petitjean}, {Prada},
  {Price-Whelan}, {Reid}, {Roe}, {Ross}, {Ross}, {Sabiu}, {Saito}, {Samushia},
  {S{\'a}nchez}, {Schlegel}, {Schneider}, {Scoccola}, {Seo}, {Skibba},
  {Strauss}, {Swanson}, {Thomas}, {Tinker}, {Tojeiro}, {Maga{\~n}a}, {Verde},
  {Wake}, {Weaver}, {Weinberg}, {White}, {Xu}, {Y{\`e}che}, {Zehavi}, \&
  {Zhao}}]{anderson2014}
{Anderson}, L., {Aubourg}, {\'E}., {Bailey}, S., {et~al.} 2014, \mnras, 441,
  24, \dodoi{10.1093/mnras/stu523}

\bibitem[{{Barriga} \& {Gazta{\~n}aga}(2002)}]{barriga2002}
{Barriga}, J., \& {Gazta{\~n}aga}, E. 2002, \mnras, 333, 443,
  \dodoi{10.1046/j.1365-8711.2002.05431.x}

\bibitem[{{Bautista} {et~al.}(2020){Bautista}, {Paviot}, {Maga{\~n}a}, {de la
  Torre}, {Fromenteau}, {Gil-Mar{\'\i}n}, {Ross}, {Burtin}, {Dawson}, {Hou},
  {Kneib}, {de Mattia}, {Percival}, {Rossi}, {Tojeiro}, {Zhao}, {Zhao}, {Alam},
  {Brownstein}, {Chapman}, {Choi}, {Chuang}, {Escoffier}, {de la Macorra}, {du
  Mas des Bourboux}, {Mohammad}, {Moon}, {M{\"u}ller}, {Nadathur}, {Newman},
  {Schneider}, {Seo}, \& {Wang}}]{bautista2020}
{Bautista}, J.~E., {Paviot}, R., {Maga{\~n}a}, M.~V., {et~al.} 2020, \mnras,
  \dodoi{10.1093/mnras/staa2800}

\bibitem[{{Bel} {et~al.}(2015){Bel}, {Hoffmann}, \& {Gazta{\~n}aga}}]{bel2015}
{Bel}, J., {Hoffmann}, K., \& {Gazta{\~n}aga}, E. 2015, \mnras, 453, 259,
  \dodoi{10.1093/mnras/stv1600}

\bibitem[{{Bernardeau} {et~al.}(2002){Bernardeau}, {Colombi}, {Gazta{\~n}aga},
  \& {Scoccimarro}}]{bernardeau2002}
{Bernardeau}, F., {Colombi}, S., {Gazta{\~n}aga}, E., \& {Scoccimarro}, R.
  2002, \physrep, 367, 1, \dodoi{10.1016/S0370-1573(02)00135-7}

\bibitem[{{Beutler} {et~al.}(2011){Beutler}, {Blake}, {Colless}, {Jones},
  {Staveley-Smith}, {Campbell}, {Parker}, {Saunders}, \&
  {Watson}}]{beutler2011}
{Beutler}, F., {Blake}, C., {Colless}, M., {et~al.} 2011, \mnras, 416, 3017,
  \dodoi{10.1111/j.1365-2966.2011.19250.x}

\bibitem[{{Blake} {et~al.}(2011){Blake}, {Kazin}, {Beutler}, {Davis},
  {Parkinson}, {Brough}, {Colless}, {Contreras}, {Couch}, {Croom}, {Croton},
  {Drinkwater}, {Forster}, {Gilbank}, {Gladders}, {Glazebrook}, {Jelliffe},
  {Jurek}, {Li}, {Madore}, {Martin}, {Pimbblet}, {Poole}, {Pracy}, {Sharp},
  {Wisnioski}, {Woods}, {Wyder}, \& {Yee}}]{blake2011}
{Blake}, C., {Kazin}, E.~A., {Beutler}, F., {et~al.} 2011, \mnras, 418, 1707,
  \dodoi{10.1111/j.1365-2966.2011.19592.x}

\bibitem[{{Cole} {et~al.}(2005){Cole}, {Percival}, {Peacock}, {Norberg},
  {Baugh}, {Frenk}, {Baldry}, {Bland-Hawthorn}, {Bridges}, {Cannon}, {Colless},
  {Collins}, {Couch}, {Cross}, {Dalton}, {Eke}, {De Propris}, {Driver},
  {Efstathiou}, {Ellis}, {Glazebrook}, {Jackson}, {Jenkins}, {Lahav}, {Lewis},
  {Lumsden}, {Maddox}, {Madgwick}, {Peterson}, {Sutherland}, \&
  {Taylor}}]{cole2005}
{Cole}, S., {Percival}, W.~J., {Peacock}, J.~A., {et~al.} 2005, \mnras, 362,
  505, \dodoi{10.1111/j.1365-2966.2005.09318.x}

\bibitem[{{Contarini} {et~al.}(2019){Contarini}, {Ronconi}, {Marulli},
  {Moscardini}, {Veropalumbo}, \& {Baldi}}]{contarini2019}
{Contarini}, S., {Ronconi}, T., {Marulli}, F., {et~al.} 2019, \mnras, 488,
  3526, \dodoi{10.1093/mnras/stz1989}

\bibitem[{{Costanzi} {et~al.}(2019){Costanzi}, {Rozo}, {Simet}, {Zhang},
  {Evrard}, {Mantz}, {Rykoff}, {Jeltema}, {Gruen}, {Allen}, {McClintock},
  {Romer}, {von der Linden}, {Farahi}, {DeRose}, {Varga}, {Weller}, {Giles},
  {Hollowood}, {Bhargava}, {Bermeo-Hernandez}, {Chen}, {Abbott}, {Abdalla},
  {Avila}, {Bechtol}, {Brooks}, {Buckley-Geer}, {Burke}, {Rosell}, {Kind},
  {Carretero}, {Crocce}, {Cunha}, {da Costa}, {Davis}, {De Vicente}, {Diehl},
  {Dietrich}, {Doel}, {Eifler}, {Estrada}, {Flaugher}, {Fosalba}, {Frieman},
  {Garc{\'\i}a-Bellido}, {Gaztanaga}, {Gerdes}, {Giannantonio}, {Gruendl},
  {Gschwend}, {Gutierrez}, {Hartley}, {Honscheid}, {Hoyle}, {James}, {Krause},
  {Kuehn}, {Kuropatkin}, {Lima}, {Lin}, {Maia}, {March}, {Marshall}, {Martini},
  {Menanteau}, {Miller}, {Miquel}, {Mohr}, {Ogando}, {Plazas}, {Roodman},
  {Sanchez}, {Scarpine}, {Schindler}, {Schubnell}, {Serrano}, {Sevilla-Noarbe},
  {Sheldon}, {Smith}, {Soares-Santos}, {Sobreira}, {Suchyta}, {Swanson},
  {Tarle}, {Thomas}, \& {Wechsler}}]{costanzi2019}
{Costanzi}, M., {Rozo}, E., {Simet}, M., {et~al.} 2019, \mnras, 488, 4779,
  \dodoi{10.1093/mnras/stz1949}

\bibitem[{{Davis} \& {Peebles}(1983)}]{davis1983}
{Davis}, M., \& {Peebles}, P.~J.~E. 1983, \apj, 267, 465,
  \dodoi{10.1086/160884}

\bibitem[{{Dawson} {et~al.}(2013){Dawson}, {Schlegel}, {Ahn}, {Anderson},
  {Aubourg}, {Bailey}, {Barkhouser}, {Bautista}, {Beifiori}, {Berlind},
  {Bhardwaj}, {Bizyaev}, {Blake}, {Blanton}, {Blomqvist}, {Bolton}, {Borde},
  {Bovy}, {Brandt}, {Brewington}, {Brinkmann}, {Brown}, {Brownstein}, {Bundy},
  {Busca}, {Carithers}, {Carnero}, {Carr}, {Chen}, {Comparat}, {Connolly},
  {Cope}, {Croft}, {Cuesta}, {da Costa}, {Davenport}, {Delubac}, {de Putter},
  {Dhital}, {Ealet}, {Ebelke}, {Eisenstein}, {Escoffier}, {Fan}, {Filiz Ak},
  {Finley}, {Font-Ribera}, {G{\'e}nova-Santos}, {Gunn}, {Guo}, {Haggard},
  {Hall}, {Hamilton}, {Harris}, {Harris}, {Ho}, {Hogg}, {Holder}, {Honscheid},
  {Huehnerhoff}, {Jordan}, {Jordan}, {Kauffmann}, {Kazin}, {Kirkby}, {Klaene},
  {Kneib}, {Le Goff}, {Lee}, {Long}, {Loomis}, {Lundgren}, {Lupton}, {Maia},
  {Makler}, {Malanushenko}, {Malanushenko}, {Mandelbaum}, {Manera}, {Maraston},
  {Margala}, {Masters}, {McBride}, {McDonald}, {McGreer}, {McMahon}, {Mena},
  {Miralda-Escud{\'e}}, {Montero-Dorta}, {Montesano}, {Muna}, {Myers},
  {Naugle}, {Nichol}, {Noterdaeme}, {Nuza}, {Olmstead}, {Oravetz}, {Oravetz},
  {Owen}, {Padmanabhan}, {Palanque-Delabrouille}, {Pan}, {Parejko},
  {P{\^a}ris}, {Percival}, {P{\'e}rez-Fournon}, {P{\'e}rez-R{\`a}fols},
  {Petitjean}, {Pfaffenberger}, {Pforr}, {Pieri}, {Prada}, {Price-Whelan},
  {Raddick}, {Rebolo}, {Rich}, {Richards}, {Rockosi}, {Roe}, {Ross}, {Ross},
  {Rossi}, {Rubi{\~n}o-Martin}, {Samushia}, {S{\'a}nchez}, {Sayres}, {Schmidt},
  {Schneider}, {Sc{\'o}ccola}, {Seo}, {Shelden}, {Sheldon}, {Shen}, {Shu},
  {Slosar}, {Smee}, {Snedden}, {Stauffer}, {Steele}, {Strauss}, {Streblyanska},
  {Suzuki}, {Swanson}, {Tal}, {Tanaka}, {Thomas}, {Tinker}, {Tojeiro},
  {Tremonti}, {Vargas Maga{\~n}a}, {Verde}, {Viel}, {Wake}, {Watson}, {Weaver},
  {Weinberg}, {Weiner}, {West}, {White}, {Wood-Vasey}, {Yeche}, {Zehavi},
  {Zhao}, \& {Zheng}}]{dawson2013}
{Dawson}, K.~S., {Schlegel}, D.~J., {Ahn}, C.~P., {et~al.} 2013, \aj, 145, 10,
  \dodoi{10.1088/0004-6256/145/1/10}

\bibitem[{{de Carvalho} {et~al.}(2020){de Carvalho}, {Bernui}, {Xavier}, \&
  {Novaes}}]{decarvalho2020}
{de Carvalho}, E., {Bernui}, A., {Xavier}, H.~S., \& {Novaes}, C.~P. 2020,
  \mnras, 492, 4469, \dodoi{10.1093/mnras/staa119}

\bibitem[{{Desjacques} {et~al.}(2018){Desjacques}, {Jeong}, \&
  {Schmidt}}]{desjacques2018}
{Desjacques}, V., {Jeong}, D., \& {Schmidt}, F. 2018, \physrep, 733, 1,
  \dodoi{10.1016/j.physrep.2017.12.002}

\bibitem[{{Eisenstein} \& {Hu}(1998)}]{eisenstein1998}
{Eisenstein}, D.~J., \& {Hu}, W. 1998, \apj, 496, 605, \dodoi{10.1086/305424}

\bibitem[{{Eisenstein} {et~al.}(2005){Eisenstein}, {Zehavi}, {Hogg},
  {Scoccimarro}, {Blanton}, {Nichol}, {Scranton}, {Seo}, {Tegmark}, {Zheng},
  {Anderson}, {Annis}, {Bahcall}, {Brinkmann}, {Burles}, {Castand er},
  {Connolly}, {Csabai}, {Doi}, {Fukugita}, {Frieman}, {Glazebrook}, {Gunn},
  {Hendry}, {Hennessy}, {Ivezi{\'c}}, {Kent}, {Knapp}, {Lin}, {Loh}, {Lupton},
  {Margon}, {McKay}, {Meiksin}, {Munn}, {Pope}, {Richmond}, {Schlegel},
  {Schneider}, {Shimasaku}, {Stoughton}, {Strauss}, {SubbaRao}, {Szalay},
  {Szapudi}, {Tucker}, {Yanny}, \& {York}}]{eisenstein2005}
{Eisenstein}, D.~J., {Zehavi}, I., {Hogg}, D.~W., {et~al.} 2005, \apj, 633,
  560, \dodoi{10.1086/466512}

\bibitem[{{Estrada} {et~al.}(2009){Estrada}, {Sefusatti}, \&
  {Frieman}}]{estrada2009}
{Estrada}, J., {Sefusatti}, E., \& {Frieman}, J.~A. 2009, \apj, 692, 265,
  \dodoi{10.1088/0004-637X/692/1/265}

\bibitem[{{Foreman-Mackey} {et~al.}(2013){Foreman-Mackey}, {Hogg}, {Lang}, \&
  {Goodman}}]{emcee}
{Foreman-Mackey}, D., {Hogg}, D.~W., {Lang}, D., \& {Goodman}, J. 2013, \pasp,
  125, 306, \dodoi{10.1086/670067}

\bibitem[{{Fosalba} {et~al.}(2005){Fosalba}, {Pan}, \& {Szapudi}}]{fosalba2005}
{Fosalba}, P., {Pan}, J., \& {Szapudi}, I. 2005, \apj, 632, 29,
  \dodoi{10.1086/432906}

\bibitem[{{Frieman} \& {Gaztanaga}(1994)}]{frieman1994}
{Frieman}, J.~A., \& {Gaztanaga}, E. 1994, \apj, 425, 392,
  \dodoi{10.1086/173995}

\bibitem[{{Fry}(1994)}]{fry1994}
{Fry}, J.~N. 1994, Physical Review Letters, 73, 215,
  \dodoi{10.1103/PhysRevLett.73.215}

\bibitem[{{Fry} \& {Gaztanaga}(1993)}]{fry1993}
{Fry}, J.~N., \& {Gaztanaga}, E. 1993, \apj, 413, 447, \dodoi{10.1086/173015}

\bibitem[{{Garc{\'\i}a-Farieta} {et~al.}(2020){Garc{\'\i}a-Farieta}, {Marulli},
  {Moscardini}, {Veropalumbo}, \& {Casas-Mirand a}}]{garcia2020}
{Garc{\'\i}a-Farieta}, J.~E., {Marulli}, F., {Moscardini}, L., {Veropalumbo},
  A., \& {Casas-Mirand a}, R.~A. 2020, \mnras, 494, 1658,
  \dodoi{10.1093/mnras/staa791}

\bibitem[{{Gazta{\~n}aga} {et~al.}(2009){Gazta{\~n}aga}, {Cabr{\'e}}, {Castand
  er}, {Crocce}, \& {Fosalba}}]{gaztanaga2009}
{Gazta{\~n}aga}, E., {Cabr{\'e}}, A., {Castand er}, F., {Crocce}, M., \&
  {Fosalba}, P. 2009, \mnras, 399, 801,
  \dodoi{10.1111/j.1365-2966.2009.15313.x}

\bibitem[{{Gazta{\~n}aga} \& {Scoccimarro}(2005)}]{gaztanaga2005}
{Gazta{\~n}aga}, E., \& {Scoccimarro}, R. 2005, \mnras, 361, 824,
  \dodoi{10.1111/j.1365-2966.2005.09234.x}

\bibitem[{{Gil-Mar{\'\i}n} {et~al.}(2020){Gil-Mar{\'\i}n}, {Bautista},
  {Paviot}, {Vargas-Maga{\~n}a}, {de la Torre}, {Fromenteau}, {Alam},
  {{\'A}vila}, {Burtin}, {Chuang}, {Dawson}, {Hou}, {de Mattia}, {Mohammad},
  {M{\"u}ller}, {Nadathur}, {Neveux}, {Percival}, {Raichoor}, {Rezaie}, {Ross},
  {Rossi}, {Ruhlmann-Kleider}, {Smith}, {Tamone}, {Tinker}, {Tojeiro}, {Wang},
  {Zhao}, {Zhao}, {Brinkmann}, {Brownstein}, {Choi}, {Escoffier}, {de la
  Macorra}, {Moon}, {Newman}, {Schneider}, {Seo}, \& {Vivek}}]{gilmarin2020}
{Gil-Mar{\'\i}n}, H., {Bautista}, J.~E., {Paviot}, R., {et~al.} 2020, arXiv
  e-prints, arXiv:2007.08994.
\newblock \doarXiv{2007.08994}

\bibitem[{{Goodman} \& {Weare}(2010)}]{goodman2010}
{Goodman}, J., \& {Weare}, J. 2010, Communications in Applied Mathematics and
  Computational Science, 5, 65, \dodoi{10.2140/camcos.2010.5.65}

\bibitem[{{Groth} \& {Peebles}(1977)}]{groth1977}
{Groth}, E.~J., \& {Peebles}, P.~J.~E. 1977, \apj, 217, 385,
  \dodoi{10.1086/155588}

\bibitem[{{Guo} \& {Jing}(2009)}]{guo2009}
{Guo}, H., \& {Jing}, Y.~P. 2009, \apj, 702, 425,
  \dodoi{10.1088/0004-637X/702/1/425}

\bibitem[{{Guo} {et~al.}(2014){Guo}, {Li}, {Jing}, \& {B{\"o}rner}}]{guo2014}
{Guo}, H., {Li}, C., {Jing}, Y.~P., \& {B{\"o}rner}, G. 2014, \apj, 780, 139,
  \dodoi{10.1088/0004-637X/780/2/139}

\bibitem[{{Guo} {et~al.}(2015){Guo}, {Zheng}, {Jing}, {et~al.}}]{guo2015}
{Guo}, H., {Zheng}, Z., {Jing}, Y.~P., {et~al.} 2015, \mnras, 449, L95,
  \dodoi{10.1093/mnrasl/slv020}

\bibitem[{{Guo} {et~al.}(2016){Guo}, {Zheng}, {Behroozi}, {Zehavi}, {Comparat},
  {Favole}, {Gottl{\"o}ber}, {Klypin}, {Prada}, {Rodr{\'\i}guez-Torres},
  {Weinberg}, \& {Yepes}}]{guo2016}
{Guo}, H., {Zheng}, Z., {Behroozi}, P.~S., {et~al.} 2016, \apj, 831, 3,
  \dodoi{10.3847/0004-637X/831/1/3}

\bibitem[{{Hamilton}(1992)}]{hamilton1992}
{Hamilton}, A.~J.~S. 1992, \apjl, 385, L5, \dodoi{10.1086/186264}

\bibitem[{{Hamilton}(1993)}]{hamilton1993}
---. 1993, \apj, 417, 19, \dodoi{10.1086/173288}

\bibitem[{{Hewett}(1982)}]{hewett1982}
{Hewett}, P.~C. 1982, \mnras, 201, 867, \dodoi{10.1093/mnras/201.4.867}

\bibitem[{{Hoffmann} {et~al.}(2015){Hoffmann}, {Bel}, {Gazta{\~n}aga},
  {Crocce}, {Fosalba}, \& {Castander}}]{hoffmann2015}
{Hoffmann}, K., {Bel}, J., {Gazta{\~n}aga}, E., {et~al.} 2015, \mnras, 447,
  1724, \dodoi{10.1093/mnras/stu2492}

\bibitem[{{Hong} {et~al.}(2016){Hong}, {Han}, \& {Wen}}]{hong2016}
{Hong}, T., {Han}, J.~L., \& {Wen}, Z.~L. 2016, \apj, 826, 154,
  \dodoi{10.3847/0004-637X/826/2/154}

\bibitem[{{Hong} {et~al.}(2012){Hong}, {Han}, {Wen}, {Sun}, \&
  {Zhan}}]{hong2012}
{Hong}, T., {Han}, J.~L., {Wen}, Z.~L., {Sun}, L., \& {Zhan}, H. 2012, \apj,
  749, 81, \dodoi{10.1088/0004-637X/749/1/81}

\bibitem[{{Huchra} \& {Geller}(1982)}]{huchra1982}
{Huchra}, J.~P., \& {Geller}, M.~J. 1982, \apj, 257, 423,
  \dodoi{10.1086/160000}

\bibitem[{{Hunter}(2007)}]{hunter2007}
{Hunter}, J.~D. 2007, Computing in Science and Engineering, 9, 90,
  \dodoi{10.1109/MCSE.2007.55}

\bibitem[{{Huterer} \& {Shafer}(2018)}]{huterer2018}
{Huterer}, D., \& {Shafer}, D.~L. 2018, Reports on Progress in Physics, 81,
  016901, \dodoi{10.1088/1361-6633/aa997e}

\bibitem[{{H{\"u}tsi}(2010)}]{hutsi2010}
{H{\"u}tsi}, G. 2010, \mnras, 401, 2477,
  \dodoi{10.1111/j.1365-2966.2009.15824.x}

\bibitem[{{Ivezi{\'c}} {et~al.}(2019){Ivezi{\'c}}, {Kahn}, {Tyson}, {Abel},
  {Acosta}, {Allsman}, {Alonso}, {AlSayyad}, {Anderson}, {Andrew}, {Angel},
  {Angeli}, {Ansari}, {Antilogus}, {Araujo}, {Armstrong}, {Arndt}, {Astier},
  {Aubourg}, {Auza}, {Axelrod}, {Bard}, {Barr}, {Barrau}, {Bartlett}, {Bauer},
  {Bauman}, {Baumont}, {Bechtol}, {Bechtol}, {Becker}, {Becla}, {Beldica},
  {Bellavia}, {Bianco}, {Biswas}, {Blanc}, {Blazek}, {Bland ford}, {Bloom},
  {Bogart}, {Bond}, {Booth}, {Borgland}, {Borne}, {Bosch}, {Boutigny},
  {Brackett}, {Bradshaw}, {Brand t}, {Brown}, {Bullock}, {Burchat}, {Burke},
  {Cagnoli}, {Calabrese}, {Callahan}, {Callen}, {Carlin}, {Carlson}, {Chand
  rasekharan}, {Charles-Emerson}, {Chesley}, {Cheu}, {Chiang}, {Chiang},
  {Chirino}, {Chow}, {Ciardi}, {Claver}, {Cohen-Tanugi}, {Cockrum}, {Coles},
  {Connolly}, {Cook}, {Cooray}, {Covey}, {Cribbs}, {Cui}, {Cutri}, {Daly},
  {Daniel}, {Daruich}, {Daubard}, {Daues}, {Dawson}, {Delgado}, {Dellapenna},
  {de Peyster}, {de Val-Borro}, {Digel}, {Doherty}, {Dubois},
  {Dubois-Felsmann}, {Durech}, {Economou}, {Eifler}, {Eracleous}, {Emmons},
  {Fausti Neto}, {Ferguson}, {Figueroa}, {Fisher-Levine}, {Focke}, {Foss},
  {Frank}, {Freemon}, {Gangler}, {Gawiser}, {Geary}, {Gee}, {Geha}, {Gessner},
  {Gibson}, {Gilmore}, {Glanzman}, {Glick}, {Goldina}, {Goldstein}, {Goodenow},
  {Graham}, {Gressler}, {Gris}, {Guy}, {Guyonnet}, {Haller}, {Harris},
  {Hascall}, {Haupt}, {Hernand ez}, {Herrmann}, {Hileman}, {Hoblitt},
  {Hodgson}, {Hogan}, {Howard}, {Huang}, {Huffer}, {Ingraham}, {Innes},
  {Jacoby}, {Jain}, {Jammes}, {Jee}, {Jenness}, {Jernigan}, {Jevremovi{\'c}},
  {Johns}, {Johnson}, {Johnson}, {Jones}, {Juramy-Gilles}, {Juri{\'c}},
  {Kalirai}, {Kallivayalil}, {Kalmbach}, {Kantor}, {Karst}, {Kasliwal},
  {Kelly}, {Kessler}, {Kinnison}, {Kirkby}, {Knox}, {Kotov}, {Krabbendam},
  {Krughoff}, {Kub{\'a}nek}, {Kuczewski}, {Kulkarni}, {Ku}, {Kurita}, {Lage},
  {Lambert}, {Lange}, {Langton}, {Le Guillou}, {Levine}, {Liang}, {Lim},
  {Lintott}, {Long}, {Lopez}, {Lotz}, {Lupton}, {Lust}, {MacArthur}, {Mahabal},
  {Mand elbaum}, {Markiewicz}, {Marsh}, {Marshall}, {Marshall}, {May},
  {McKercher}, {McQueen}, {Meyers}, {Migliore}, {Miller}, {Mills}, {Miraval},
  {Moeyens}, {Moolekamp}, {Monet}, {Moniez}, {Monkewitz}, {Montgomery},
  {Morrison}, {Mueller}, {Muller}, {Mu{\~n}oz Arancibia}, {Neill}, {Newbry},
  {Nief}, {Nomerotski}, {Nordby}, {O'Connor}, {Oliver}, {Olivier}, {Olsen},
  {O'Mullane}, {Ortiz}, {Osier}, {Owen}, {Pain}, {Palecek}, {Parejko},
  {Parsons}, {Pease}, {Peterson}, {Peterson}, {Petravick}, {Libby Petrick},
  {Petry}, {Pierfederici}, {Pietrowicz}, {Pike}, {Pinto}, {Plante}, {Plate},
  {Plutchak}, {Price}, {Prouza}, {Radeka}, {Rajagopal}, {Rasmussen},
  {Regnault}, {Reil}, {Reiss}, {Reuter}, {Ridgway}, {Riot}, {Ritz}, {Robinson},
  {Roby}, {Roodman}, {Rosing}, {Roucelle}, {Rumore}, {Russo}, {Saha},
  {Sassolas}, {Schalk}, {Schellart}, {Schindler}, {Schmidt}, {Schneider},
  {Schneider}, {Schoening}, {Schumacher}, {Schwamb}, {Sebag}, {Selvy},
  {Sembroski}, {Seppala}, {Serio}, {Serrano}, {Shaw}, {Shipsey}, {Sick},
  {Silvestri}, {Slater}, {Smith}, {Smith}, {Sobhani}, {Soldahl},
  {Storrie-Lombardi}, {Stover}, {Strauss}, {Street}, {Stubbs}, {Sullivan},
  {Sweeney}, {Swinbank}, {Szalay}, {Takacs}, {Tether}, {Thaler}, {Thayer},
  {Thomas}, {Thornton}, {Thukral}, {Tice}, {Trilling}, {Turri}, {Van Berg},
  {Vanden Berk}, {Vetter}, {Virieux}, {Vucina}, {Wahl}, {Walkowicz}, {Walsh},
  {Walter}, {Wang}, {Wang}, {Warner}, {Wiecha}, {Willman}, {Winters},
  {Wittman}, {Wolff}, {Wood-Vasey}, {Wu}, {Xin}, {Yoachim}, \&
  {Zhan}}]{lsst2019}
{Ivezi{\'c}}, {\v{Z}}., {Kahn}, S.~M., {Tyson}, J.~A., {et~al.} 2019, \apj,
  873, 111, \dodoi{10.3847/1538-4357/ab042c}

\bibitem[{{Jing} \& {B{\"o}rner}(1998)}]{jing1998}
{Jing}, Y.~P., \& {B{\"o}rner}, G. 1998, \apj, 503, 37, \dodoi{10.1086/305997}

\bibitem[{{Jing} \& {B{\"o}rner}(2004)}]{jing2004}
---. 2004, \apj, 607, 140, \dodoi{10.1086/383343}

\bibitem[{{Jing} {et~al.}(1995){Jing}, {Borner}, \& {Valdarnini}}]{jing1995}
{Jing}, Y.~P., {Borner}, G., \& {Valdarnini}, R. 1995, \mnras, 277, 630

\bibitem[{{Kaiser}(1987)}]{kaiser1987}
{Kaiser}, N. 1987, \mnras, 227, 1

\bibitem[{{Kayo} {et~al.}(2004){Kayo}, {Suto}, {Nichol}, {Pan}, {Szapudi},
  {Connolly}, {Gardner}, {Jain}, {Kulkarni}, {Matsubara}, {Sheth}, {Szalay}, \&
  {Brinkmann}}]{kayo2004}
{Kayo}, I., {Suto}, Y., {Nichol}, R.~C., {et~al.} 2004, \pasj, 56, 415,
  \dodoi{10.1093/pasj/56.3.415}

\bibitem[{{Keih{\"a}nen} {et~al.}(2019){Keih{\"a}nen}, {Kurki-Suonio},
  {Lindholm}, {Viitanen}, {Suur-Uski}, {Allevato}, {Branchini}, {Marulli},
  {Norberg}, {Tavagnacco}, {de la Torre}, {Valiviita}, {Viel}, {Bel},
  {Frailis}, \& {S{\'a}nchez}}]{keihanen2019}
{Keih{\"a}nen}, E., {Kurki-Suonio}, H., {Lindholm}, V., {et~al.} 2019, \aap,
  631, A73, \dodoi{10.1051/0004-6361/201935828}

\bibitem[{{Kulkarni} {et~al.}(2007){Kulkarni}, {Nichol}, {Sheth},
  {et~al.}}]{kulkarni2007}
{Kulkarni}, G.~V., {Nichol}, R.~C., {Sheth}, R.~K., {et~al.} 2007, \mnras, 378,
  1196, \dodoi{10.1111/j.1365-2966.2007.11872.x}

\bibitem[{{Landy} \& {Szalay}(1993)}]{landy1993}
{Landy}, S.~D., \& {Szalay}, A.~S. 1993, \apj, 412, 64, \dodoi{10.1086/172900}

\bibitem[{{Laureijs} {et~al.}(2011){Laureijs}, {Amiaux}, {Arduini},
  {Augu{\`e}res}, {Brinchmann}, {Cole}, {Cropper}, {Dabin}, {Duvet}, {Ealet},
  \& et~al.}]{euclid}
{Laureijs}, R., {Amiaux}, J., {Arduini}, S., {et~al.} 2011, ArXiv e-prints.
\newblock \doarXiv{1110.3193}

\bibitem[{{Lazeyras} {et~al.}(2016){Lazeyras}, {Wagner}, {Baldauf}, \&
  {Schmidt}}]{lazeyras2016}
{Lazeyras}, T., {Wagner}, C., {Baldauf}, T., \& {Schmidt}, F. 2016, \jcap,
  2016, 018, \dodoi{10.1088/1475-7516/2016/02/018}

\bibitem[{{Lesci} {et~al.}(in prep.)}]{lesci2020}
{Lesci}, G., {et~al.} in prep.

\bibitem[{{Lewis} \& {Bridle}(2002)}]{lewis2002}
{Lewis}, A., \& {Bridle}, S. 2002, \prd, 66, 103511,
  \dodoi{10.1103/PhysRevD.66.103511}

\bibitem[{{Lewis} {et~al.}(2000){Lewis}, {Challinor}, \& {Lasenby}}]{lewis2000}
{Lewis}, A., {Challinor}, A., \& {Lasenby}, A. 2000, \apj, 538, 473,
  \dodoi{10.1086/309179}

\bibitem[{{Maraston} {et~al.}(2013){Maraston}, {Pforr}, {Henriques}, {Thomas},
  {Wake}, {Brownstein}, {Capozzi}, {Tinker}, {Bundy}, {Skibba}, {Beifiori},
  {Nichol}, {Edmondson}, {Schneider}, {Chen}, {Masters}, {Steele}, {Bolton},
  {York}, {Weaver}, {Higgs}, {Bizyaev}, {Brewington}, {Malanushenko},
  {Malanushenko}, {Snedden}, {Oravetz}, {Pan}, {Shelden}, \&
  {Simmons}}]{maraston2013}
{Maraston}, C., {Pforr}, J., {Henriques}, B.~M., {et~al.} 2013, \mnras, 435,
  2764, \dodoi{10.1093/mnras/stt1424}

\bibitem[{{Mar{\'{\i}}n}(2011)}]{marin2011}
{Mar{\'{\i}}n}, F. 2011, \apj, 737, 97, \dodoi{10.1088/0004-637X/737/2/97}

\bibitem[{{Mar{\'\i}n} {et~al.}(2008){Mar{\'\i}n}, {Wechsler}, {Frieman}, \&
  {Nichol}}]{marin2008}
{Mar{\'\i}n}, F.~A., {Wechsler}, R.~H., {Frieman}, J.~A., \& {Nichol}, R.~C.
  2008, \apj, 672, 849, \dodoi{10.1086/523628}

\bibitem[{{Mar{\'\i}n} {et~al.}(2013){Mar{\'\i}n}, {Blake}, {Poole}, {McBride},
  {Brough}, {Colless}, {Contreras}, {Couch}, {Croton}, {Croom}, {Davis},
  {Drinkwater}, {Forster}, {Gilbank}, {Gladders}, {Glazebrook}, {Jelliffe},
  {Jurek}, {Li}, {Madore}, {Martin}, {Pimbblet}, {Pracy}, {Sharp}, {Wisnioski},
  {Woods}, {Wyder}, \& {Yee}}]{marin2013}
{Mar{\'\i}n}, F.~A., {Blake}, C., {Poole}, G.~B., {et~al.} 2013, \mnras, 432,
  2654, \dodoi{10.1093/mnras/stt520}

\bibitem[{{Marulli} {et~al.}(2020){Marulli}, {Veropalumbo},
  {Garc{\'\i}a-Farieta}, {Moresco}, {Moscardini}, \& {Cimatti}}]{marulli2020}
{Marulli}, F., {Veropalumbo}, A., {Garc{\'\i}a-Farieta}, J.~E., {et~al.} 2020,
  arXiv e-prints, arXiv:2010.11206.
\newblock \doarXiv{2010.11206}

\bibitem[{{Marulli} {et~al.}(2016){Marulli}, {Veropalumbo}, \& {Moresco}}]{CBL}
{Marulli}, F., {Veropalumbo}, A., \& {Moresco}, M. 2016, Astronomy and
  Computing, 14, 35, \dodoi{10.1016/j.ascom.2016.01.005}

\bibitem[{{Marulli} {et~al.}(2017){Marulli}, {Veropalumbo}, {Moscardini},
  {Cimatti}, \& {Dolag}}]{marulli2017}
{Marulli}, F., {Veropalumbo}, A., {Moscardini}, L., {Cimatti}, A., \& {Dolag},
  K. 2017, \aap, 599, A106, \dodoi{10.1051/0004-6361/201526885}

\bibitem[{{Marulli} {et~al.}(2018){Marulli}, {Veropalumbo}, {Sereno},
  {Moscardini}, {Pacaud}, {Pierre}, {Plionis}, {Cappi}, {Adami}, {Alis},
  {Altieri}, {Birkinshaw}, {Ettori}, {Faccioli}, {Gastaldello}, {Koulouridis},
  {Lidman}, {Le F{\`e}vre}, {Maurogordato}, {Poggianti}, {Pompei},
  {Sadibekova}, \& {Valtchanov}}]{marulli2018}
{Marulli}, F., {Veropalumbo}, A., {Sereno}, M., {et~al.} 2018, \aap, 620, A1,
  \dodoi{10.1051/0004-6361/201833238}

\bibitem[{{McBride} {et~al.}(2011){McBride}, {Connolly}, {Gardner}, {Scranton},
  {Scoccimarro}, {Berlind}, {Mar{\'\i}n}, \& {Schneider}}]{mcbride2011}
{McBride}, C.~K., {Connolly}, A.~J., {Gardner}, J.~P., {et~al.} 2011, \apj,
  739, 85, \dodoi{10.1088/0004-637X/739/2/85}

\bibitem[{{Moresco} {et~al.}(2014){Moresco}, {Marulli}, {Baldi}, {Moscardini},
  \& {Cimatti}}]{moresco2014}
{Moresco}, M., {Marulli}, F., {Baldi}, M., {Moscardini}, L., \& {Cimatti}, A.
  2014, \mnras, 443, 2874, \dodoi{10.1093/mnras/stu1359}

\bibitem[{{Moresco} {et~al.}(2017){Moresco}, {Marulli}, {Moscardini},
  {Branchini}, {Cappi}, {Davidzon}, {Granett}, {de la Torre}, {Guzzo}, {Abbas},
  {Adami}, {Arnouts}, {Bel}, {Bolzonella}, {Bottini}, {Carbone}, {Coupon},
  {Cucciati}, {De Lucia}, {Franzetti}, {Fritz}, {Fumana}, {Garilli}, {Ilbert},
  {Iovino}, {Krywult}, {Le Brun}, {Le F{\`e}vre}, {Ma{\l}ek}, {McCracken},
  {Polletta}, {Pollo}, {Scodeggio}, {Tasca}, {Tojeiro}, {Vergani}, \&
  {Zanichelli}}]{moresco2017}
{Moresco}, M., {Marulli}, F., {Moscardini}, L., {et~al.} 2017, \aap, 604, A133,
  \dodoi{10.1051/0004-6361/201628589}

\bibitem[{{Nanni} {et~al.}(in prep.)}]{nanni2020}
{Nanni}, L., {et~al.} in prep.

\bibitem[{{Nichol} {et~al.}(2006){Nichol}, {Sheth}, {Suto},
  {et~al.}}]{nichol2006}
{Nichol}, R.~C., {Sheth}, R.~K., {Suto}, Y., {et~al.} 2006, \mnras, 368, 1507,
  \dodoi{10.1111/j.1365-2966.2006.10239.x}

\bibitem[{{Pacaud} {et~al.}(2018){Pacaud}, {Pierre}, {Melin}, {Adami},
  {Evrard}, {Galli}, {Gastaldello}, {Maughan}, {Sereno}, {Alis}, {Altieri},
  {Birkinshaw}, {Chiappetti}, {Faccioli}, {Giles}, {Horellou}, {Iovino},
  {Koulouridis}, {Le F{\`e}vre}, {Lidman}, {Lieu}, {Maurogordato},
  {Moscardini}, {Plionis}, {Poggianti}, {Pompei}, {Sadibekova}, {Valtchanov},
  \& {Willis}}]{pacaud2018}
{Pacaud}, F., {Pierre}, M., {Melin}, J.~B., {et~al.} 2018, \aap, 620, A10,
  \dodoi{10.1051/0004-6361/201834022}

\bibitem[{{Padmanabhan} {et~al.}(2012){Padmanabhan}, {Xu}, {Eisenstein},
  {Scalzo}, {Cuesta}, {Mehta}, \& {Kazin}}]{padmanabhan2012}
{Padmanabhan}, N., {Xu}, X., {Eisenstein}, D.~J., {et~al.} 2012, \mnras, 427,
  2132, \dodoi{10.1111/j.1365-2966.2012.21888.x}

\bibitem[{{Peebles}(1980)}]{peebles1980}
{Peebles}, P.~J.~E. 1980, {The large-scale structure of the universe}

\bibitem[{{Peebles} \& {Groth}(1975)}]{peebles1975}
{Peebles}, P.~J.~E., \& {Groth}, E.~J. 1975, \apj, 196, 1,
  \dodoi{10.1086/153390}

\bibitem[{{Peebles} \& {Hauser}(1974)}]{peebles1974}
{Peebles}, P.~J.~E., \& {Hauser}, M.~G. 1974, \apjs, 28, 19,
  \dodoi{10.1086/190308}

\bibitem[{{Peebles} \& {Yu}(1970)}]{peebles1970}
{Peebles}, P.~J.~E., \& {Yu}, J.~T. 1970, \apj, 162, 815,
  \dodoi{10.1086/150713}

\bibitem[{{Percival} {et~al.}(2007){Percival}, {Cole}, {Eisenstein}, {Nichol},
  {Peacock}, {Pope}, \& {Szalay}}]{percival2007}
{Percival}, W.~J., {Cole}, S., {Eisenstein}, D.~J., {et~al.} 2007, \mnras, 381,
  1053, \dodoi{10.1111/j.1365-2966.2007.12268.x}

\bibitem[{{Perlmutter} {et~al.}(1999){Perlmutter}, {Aldering}, {Goldhaber},
  {Knop}, {Nugent}, {Castro}, {Deustua}, {Fabbro}, {Goobar}, {Groom}, {Hook},
  {Kim}, {Kim}, {Lee}, {Nunes}, {Pain}, {Pennypacker}, {Quimby}, {Lidman},
  {Ellis}, {Irwin}, {McMahon}, {Ruiz-Lapuente}, {Walton}, {Schaefer}, {Boyle},
  {Filippenko}, {Matheson}, {Fruchter}, {Panagia}, {Newberg}, {Couch}, \&
  {Project}}]{perlmutter1999}
{Perlmutter}, S., {Aldering}, G., {Goldhaber}, G., {et~al.} 1999, \apj, 517,
  565, \dodoi{10.1086/307221}

\bibitem[{{Pillepich} {et~al.}(2012){Pillepich}, {Porciani}, \&
  {Reiprich}}]{pillepich2012}
{Pillepich}, A., {Porciani}, C., \& {Reiprich}, T.~H. 2012, \mnras, 422, 44,
  \dodoi{10.1111/j.1365-2966.2012.20443.x}

\bibitem[{{Planck Collaboration} {et~al.}(2016){Planck Collaboration}, {Ade},
  {Aghanim}, {Arnaud}, {Ashdown}, {Aumont}, {Baccigalupi}, {Banday},
  {Barreiro}, {Bartlett}, \& et~al.}]{planck2016}
{Planck Collaboration}, {Ade}, P.~A.~R., {Aghanim}, N., {et~al.} 2016, \aap,
  594, A13, \dodoi{10.1051/0004-6361/201525830}

\bibitem[{{Planck Collaboration} {et~al.}(2018){Planck Collaboration},
  {Aghanim}, {Akrami}, {Ashdown}, {Aumont}, {Baccigalupi}, {Ballardini},
  {Banday}, {Barreiro}, {Bartolo}, {Basak}, {Battye}, {Benabed}, {Bernard},
  {Bersanelli}, {Bielewicz}, {Bock}, {Bond}, {Borrill}, {Bouchet}, {Boulanger},
  {Bucher}, {Burigana}, {Butler}, {Calabrese}, {Cardoso}, {Carron},
  {Challinor}, {Chiang}, {Chluba}, {Colombo}, {Combet}, {Contreras}, {Crill},
  {Cuttaia}, {de Bernardis}, {de Zotti}, {Delabrouille}, {Delouis}, {Di
  Valentino}, {Diego}, {Dor{\'e}}, {Douspis}, {Ducout}, {Dupac}, {Dusini},
  {Efstathiou}, {Elsner}, {En{\ss}lin}, {Eriksen}, {Fantaye}, {Farhang},
  {Fergusson}, {Fernandez-Cobos}, {Finelli}, {Forastieri}, {Frailis},
  {Fraisse}, {Franceschi}, {Frolov}, {Galeotta}, {Galli}, {Ganga},
  {G{\'e}nova-Santos}, {Gerbino}, {Ghosh}, {Gonz{\'a}lez-Nuevo}, {G{\'o}rski},
  {Gratton}, {Gruppuso}, {Gudmundsson}, {Hamann}, {Handley}, {Hansen},
  {Herranz}, {Hildebrandt}, {Hivon}, {Huang}, {Jaffe}, {Jones}, {Karakci},
  {Keih{\"a}nen}, {Keskitalo}, {Kiiveri}, {Kim}, {Kisner}, {Knox},
  {Krachmalnicoff}, {Kunz}, {Kurki-Suonio}, {Lagache}, {Lamarre}, {Lasenby},
  {Lattanzi}, {Lawrence}, {Le Jeune}, {Lemos}, {Lesgourgues}, {Levrier},
  {Lewis}, {Liguori}, {Lilje}, {Lilley}, {Lindholm}, {L{\'o}pez-Caniego},
  {Lubin}, {Ma}, {Mac{\'\i}as-P{\'e}rez}, {Maggio}, {Maino}, {Mandolesi},
  {Mangilli}, {Marcos-Caballero}, {Maris}, {Martin}, {Martinelli},
  {Mart{\'\i}nez-Gonz{\'a}lez}, {Matarrese}, {Mauri}, {McEwen}, {Meinhold},
  {Melchiorri}, {Mennella}, {Migliaccio}, {Millea}, {Mitra},
  {Miville-Desch{\^e}nes}, {Molinari}, {Montier}, {Morgante}, {Moss}, {Natoli},
  {N{\o}rgaard-Nielsen}, {Pagano}, {Paoletti}, {Partridge}, {Patanchon},
  {Peiris}, {Perrotta}, {Pettorino}, {Piacentini}, {Polastri}, {Polenta},
  {Puget}, {Rachen}, {Reinecke}, {Remazeilles}, {Renzi}, {Rocha}, {Rosset},
  {Roudier}, {Rubi{\~n}o-Mart{\'\i}n}, {Ruiz-Granados}, {Salvati}, {Sandri},
  {Savelainen}, {Scott}, {Shellard}, {Sirignano}, {Sirri}, {Spencer},
  {Sunyaev}, {Suur-Uski}, {Tauber}, {Tavagnacco}, {Tenti}, {Toffolatti},
  {Tomasi}, {Trombetti}, {Valenziano}, {Valiviita}, {Van Tent}, {Vibert},
  {Vielva}, {Villa}, {Vittorio}, {Wand elt}, {Wehus}, {White}, {White},
  {Zacchei}, \& {Zonca}}]{planck2018}
{Planck Collaboration}, {Aghanim}, N., {Akrami}, Y., {et~al.} 2018, arXiv
  e-prints, arXiv:1807.06209.
\newblock \doarXiv{1807.06209}

\bibitem[{{Reid} {et~al.}(2016){Reid}, {Ho}, {Padmanabhan}, {Percival},
  {Tinker}, {Tojeiro}, {White}, {Eisenstein}, {Maraston}, {Ross},
  {S{\'a}nchez}, {Schlegel}, {Sheldon}, {Strauss}, {Thomas}, {Wake}, {Beutler},
  {Bizyaev}, {Bolton}, {Brownstein}, {Chuang}, {Dawson}, {Harding}, {Kitaura},
  {Leauthaud}, {Masters}, {McBride}, {More}, {Olmstead}, {Oravetz}, {Nuza},
  {Pan}, {Parejko}, {Pforr}, {Prada}, {Rodr{\'\i}guez-Torres},
  {Salazar-Albornoz}, {Samushia}, {Schneider}, {Sc{\'o}ccola}, {Simmons}, \&
  {Vargas-Magana}}]{reid2016}
{Reid}, B., {Ho}, S., {Padmanabhan}, N., {et~al.} 2016, \mnras, 455, 1553,
  \dodoi{10.1093/mnras/stv2382}

\bibitem[{{Riess} {et~al.}(1998){Riess}, {Filippenko}, {Challis},
  {Clocchiatti}, {Diercks}, {Garnavich}, {Gilliland}, {Hogan}, {Jha},
  {Kirshner}, {Leibundgut}, {Phillips}, {Reiss}, {Schmidt}, {Schommer},
  {Smith}, {Spyromilio}, {Stubbs}, {Suntzeff}, \& {Tonry}}]{riess1998}
{Riess}, A.~G., {Filippenko}, A.~V., {Challis}, P., {et~al.} 1998, \aj, 116,
  1009, \dodoi{10.1086/300499}

\bibitem[{{Ronconi} {et~al.}(2019){Ronconi}, {Contarini}, {Marulli}, {Baldi},
  \& {Moscardini}}]{ronconi2019}
{Ronconi}, T., {Contarini}, S., {Marulli}, F., {Baldi}, M., \& {Moscardini}, L.
  2019, \mnras, 488, 5075, \dodoi{10.1093/mnras/stz2115}

\bibitem[{{Ronconi} \& {Marulli}(2017)}]{ronconi2017}
{Ronconi}, T., \& {Marulli}, F. 2017, \aap, 607, A24,
  \dodoi{10.1051/0004-6361/201730852}

\bibitem[{{Ross} {et~al.}(2015){Ross}, {Samushia}, {Howlett}, {Percival},
  {Burden}, \& {Manera}}]{ross2015}
{Ross}, A.~J., {Samushia}, L., {Howlett}, C., {et~al.} 2015, \mnras, 449, 835,
  \dodoi{10.1093/mnras/stv154}

\bibitem[{{Sartoris} {et~al.}(2016){Sartoris}, {Biviano}, {Fedeli}, {Bartlett},
  {Borgani}, {Costanzi}, {Giocoli}, {Moscardini}, {Weller}, {Ascaso},
  {Bardelli}, {Maurogordato}, \& {Viana}}]{sartoris2016}
{Sartoris}, B., {Biviano}, A., {Fedeli}, C., {et~al.} 2016, \mnras, 459, 1764,
  \dodoi{10.1093/mnras/stw630}

\bibitem[{{Sereno} {et~al.}(2015){Sereno}, {Veropalumbo}, {Marulli}, {Covone},
  {Moscardini}, \& {Cimatti}}]{sereno2015}
{Sereno}, M., {Veropalumbo}, A., {Marulli}, F., {et~al.} 2015, \mnras, 449,
  4147, \dodoi{10.1093/mnras/stv280}

\bibitem[{{Slepian} \& {Eisenstein}(2015)}]{slepian2015}
{Slepian}, Z., \& {Eisenstein}, D.~J. 2015, \mnras, 454, 4142,
  \dodoi{10.1093/mnras/stv2119}

\bibitem[{{Slepian} {et~al.}(2017{\natexlab{a}}){Slepian}, {Eisenstein},
  {Beutler}, {Chuang}, {Cuesta}, {Ge}, {Gil-Mar{\'\i}n}, {Ho}, {Kitaura},
  {McBride}, {Nichol}, {Percival}, {Rodr{\'\i}guez-Torres}, {Ross},
  {Scoccimarro}, {Seo}, {Tinker}, {Tojeiro}, \&
  {Vargas-Maga{\~n}a}}]{slepian2017}
{Slepian}, Z., {Eisenstein}, D.~J., {Beutler}, F., {et~al.} 2017{\natexlab{a}},
  \mnras, 468, 1070, \dodoi{10.1093/mnras/stw3234}

\bibitem[{{Slepian} {et~al.}(2017{\natexlab{b}}){Slepian}, {Eisenstein},
  {Brownstein}, {Chuang}, {Gil-Mar{\'\i}n}, {Ho}, {Kitaura}, {Percival},
  {Ross}, {Rossi}, {Seo}, {Slosar}, \& {Vargas-Maga{\~n}a}}]{slepian2017b}
{Slepian}, Z., {Eisenstein}, D.~J., {Brownstein}, J.~R., {et~al.}
  2017{\natexlab{b}}, \mnras, 469, 1738, \dodoi{10.1093/mnras/stx488}

\bibitem[{{Sosa Nu{\~n}ez} \& {Niz}(2020)}]{sosa2020}
{Sosa Nu{\~n}ez}, F., \& {Niz}, G. 2020, arXiv e-prints, arXiv:2006.05434.
\newblock \doarXiv{2006.05434}

\bibitem[{{Spergel} {et~al.}(2015){Spergel}, {Gehrels}, {Baltay}, {Bennett},
  {Breckinridge}, {Donahue}, {Dressler}, {Gaudi}, {Greene}, {Guyon}, {Hirata},
  {Kalirai}, {Kasdin}, {Macintosh}, {Moos}, {Perlmutter}, {Postman},
  {Rauscher}, {Rhodes}, {Wang}, {Weinberg}, {Benford}, {Hudson}, {Jeong},
  {Mellier}, {Traub}, {Yamada}, {Capak}, {Colbert}, {Masters}, {Penny},
  {Savransky}, {Stern}, {Zimmerman}, {Barry}, {Bartusek}, {Carpenter}, {Cheng},
  {Content}, {Dekens}, {Demers}, {Grady}, {Jackson}, {Kuan}, {Kruk}, {Melton},
  {Nemati}, {Parvin}, {Poberezhskiy}, {Peddie}, {Ruffa}, {Wallace}, {Whipple},
  {Wollack}, \& {Zhao}}]{spergel2015}
{Spergel}, D., {Gehrels}, N., {Baltay}, C., {et~al.} 2015, arXiv e-prints,
  arXiv:1503.03757.
\newblock \doarXiv{1503.03757}

\bibitem[{{Sunyaev} \& {Zeldovich}(1970)}]{sunyaev1970}
{Sunyaev}, R.~A., \& {Zeldovich}, Y.~B. 1970, \apss, 7, 3,
  \dodoi{10.1007/BF00653471}

\bibitem[{{Swanson} {et~al.}(2008){Swanson}, {Tegmark}, {Hamilton}, \&
  {Hill}}]{swanson2008}
{Swanson}, M.~E.~C., {Tegmark}, M., {Hamilton}, A. J.~S., \& {Hill}, J.~C.
  2008, \mnras, 387, 1391, \dodoi{10.1111/j.1365-2966.2008.13296.x}

\bibitem[{{Szapudi} \& {Szalay}(1998)}]{szapudi1998}
{Szapudi}, I., \& {Szalay}, A.~S. 1998, \apjl, 494, L41, \dodoi{10.1086/311146}

\bibitem[{{Takada} \& {Jain}(2003)}]{takada2003}
{Takada}, M., \& {Jain}, B. 2003, \mnras, 340, 580,
  \dodoi{10.1046/j.1365-8711.2003.06321.x}

\bibitem[{{Tinker} {et~al.}(2010){Tinker}, {Robertson}, {Kravtsov}, {Klypin},
  {Warren}, {Yepes}, \& {Gottl{\"o}ber}}]{tinker2010}
{Tinker}, J.~L., {Robertson}, B.~E., {Kravtsov}, A.~V., {et~al.} 2010, \apj,
  724, 878, \dodoi{10.1088/0004-637X/724/2/878}

\bibitem[{{Valageas} \& {Clerc}(2012)}]{valageas2012}
{Valageas}, P., \& {Clerc}, N. 2012, \aap, 547, A100,
  \dodoi{10.1051/0004-6361/201219646}

\bibitem[{{Veropalumbo} {et~al.}(2014){Veropalumbo}, {Marulli}, {Moscardini},
  {Moresco}, \& {Cimatti}}]{veropalumbo2014}
{Veropalumbo}, A., {Marulli}, F., {Moscardini}, L., {Moresco}, M., \&
  {Cimatti}, A. 2014, \mnras, 442, 3275, \dodoi{10.1093/mnras/stu1050}

\bibitem[{{Veropalumbo} {et~al.}(2016){Veropalumbo}, {Marulli}, {Moscardini},
  {Moresco}, \& {Cimatti}}]{veropalumbo2016}
---. 2016, \mnras, 458, 1909, \dodoi{10.1093/mnras/stw306}

\bibitem[{{Veropalumbo} {et~al.}(in prep.)}]{veropalumbo2020}
{Veropalumbo}, A., {et~al.} in prep.

\bibitem[{{Vikhlinin} {et~al.}(2009){Vikhlinin}, {Kravtsov}, {Burenin},
  {Ebeling}, {Forman}, {Hornstrup}, {Jones}, {Murray}, {Nagai}, {Quintana}, \&
  {Voevodkin}}]{vikhlinin2009}
{Vikhlinin}, A., {Kravtsov}, A.~V., {Burenin}, R.~A., {et~al.} 2009, \apj, 692,
  1060, \dodoi{10.1088/0004-637X/692/2/1060}

\bibitem[{{Virtanen} {et~al.}(2020){Virtanen}, {Gommers}, {Oliphant},
  {Haberland}, {Reddy}, {Cournapeau}, {Burovski}, {Peterson}, {Weckesser},
  {Bright}, {van der Walt}, {Brett}, {Wilson}, {Jarrod Millman}, {Mayorov},
  {Nelson}, {Jones}, {Kern}, {Larson}, {Carey}, {Polat}, {Feng}, {Moore}, {Vand
  erPlas}, {Laxalde}, {Perktold}, {Cimrman}, {Henriksen}, {Quintero}, {Harris},
  {Archibald}, {Ribeiro}, {Pedregosa}, {van Mulbregt}, \&
  {Contributors}}]{scipy}
{Virtanen}, P., {Gommers}, R., {Oliphant}, T.~E., {et~al.} 2020, Nature
  Methods, 17, 261, \dodoi{https://doi.org/10.1038/s41592-019-0686-2}

\bibitem[{{Wang} {et~al.}(2004){Wang}, {Yang}, {Mo}, {van den Bosch}, \&
  {Chu}}]{wang2004}
{Wang}, Y., {Yang}, X., {Mo}, H.~J., {van den Bosch}, F.~C., \& {Chu}, Y. 2004,
  \mnras, 353, 287, \dodoi{10.1111/j.1365-2966.2004.08141.x}

\bibitem[{{Wen} {et~al.}(2012){Wen}, {Han}, \& {Liu}}]{wen2012}
{Wen}, Z.~L., {Han}, J.~L., \& {Liu}, F.~S. 2012, \apjs, 199, 34,
  \dodoi{10.1088/0067-0049/199/2/34}

\bibitem[{{Zheng}(2004)}]{zheng2004}
{Zheng}, Z. 2004, \apj, 614, 527, \dodoi{10.1086/423838}

\end{thebibliography}
\bibliographystyle{aasjournal}

\end{document}